\ifdefined\Options%
\else%
    \def\Options{opta}
\fi%

\pdfoutput=1
\documentclass[twocolumn]{IEEEtran}

\usepackage{comment}
\usepackage{pgfplots}
\pgfplotsset{compat=newest}
\usetikzlibrary{plotmarks}
\usetikzlibrary{arrows.meta}
\usetikzlibrary{positioning}
\usepgfplotslibrary{patchplots}
\usepackage{grffile}
\usepackage{amsmath}
\usepackage{breqn}
\usepackage{flafter} 
\usepackage{placeins}

\usepackage{siunitx}
\usepackage[utf8]{inputenc}
\usepackage{graphicx}
\usepackage[english]{babel}
\usepackage{eqnarray}

\usepackage{multirow}
\usepackage[noadjust]{cite} 
\usepackage{arydshln}
\usepackage{longtable}
\usepackage{multirow}
\usepackage{array}

\usepackage{soul}
\usepackage{url}
\newcolumntype{M}[1]{>{\centering\arraybackslash}m{#1}}
\newcolumntype{N}{@{}m{0pt}@{}}

\usepackage{amssymb}
\usepackage{color}
\usepackage{textcomp}

\usepackage[\Options]{optional}
\usepackage{tikz}
\usepgfplotslibrary{external} 
\usepackage{subfigure}
\usepackage[percent]{overpic}
\usepackage{float}
\usepackage{algorithm}
\usepackage{algpseudocode}

\makeatletter
\renewcommand{\ALG@beginalgorithmic}
\makeatother

\errorcontextlines\maxdimen
\makeatletter
\newcommand*{\algrule}[1][\algorithmicindent]{\makebox[#1][l]{\hspace*{.5em}\thealgruleextra\vrule height \thealgruleheight depth \thealgruledepth}}%
\newcommand*{\thealgruleextra}{}
\newcommand*{\thealgruleheight}{1\baselineskip}
\newcommand*{\thealgruledepth}{.3\baselineskip}

\newcount\ALG@printindent@tempcnta
\def\ALG@printindent{%
    \ifnum \theALG@nested>0
        \ifx\ALG@text\ALG@x@notext
        \else
            \unskip
            \addvspace{-1pt}
            \ALG@printindent@tempcnta=1
            \loop
                \algrule[\csname ALG@ind@\the\ALG@printindent@tempcnta\endcsname]%
                \advance \ALG@printindent@tempcnta 1
            \ifnum \ALG@printindent@tempcnta<\numexpr\theALG@nested+1\relax
            \repeat
        \fi
    \fi
    }%
    
    \usepackage{etoolbox}
\patchcmd{\ALG@doentity}{\noindent\hskip\ALG@tlm}{\ALG@printindent}{}{\errmessage{failed to patch}}
\makeatother

\newbox\statebox
\newcommand{\myState}[1]{%
    \setbox\statebox=\vbox{#1}%
    \edef\thealgruleheight{\dimexpr \the\ht\statebox+1pt\relax}%
    \edef\thealgruledepth{\dimexpr \the\dp\statebox+1pt\relax}%
    \ifdim\thealgruleheight<.75\baselineskip
        \def\thealgruleheight{\dimexpr .75\baselineskip+1pt\relax}%
    \fi
    \ifdim\thealgruledepth<.25\baselineskip
        \def\thealgruledepth{\dimexpr .25\baselineskip+1pt\relax}%
    \fi
    \State #1%
    \def\thealgruleheight{\dimexpr .75\baselineskip+1pt\relax}%
    \def\thealgruledepth{\dimexpr .25\baselineskip+1pt\relax}%
}
\newcommand\longvdots[1]{\raisebox{1em}{\rotatebox{-90}{\hbox to #1 {\dotfill}}}}              

\usepackage{graphicx}
\usepackage[english]{babel}
\usepackage{eqnarray,amsmath}
\usepackage{multirow}
\usepackage[noadjust]{cite} 
\usepackage{siunitx}
\usepackage[cuteinductors,smartlabels]{circuitikz}

\usetikzlibrary{calc}
\ctikzset{bipoles/thickness=0.5}
\ctikzset{bipoles/length=0.8cm}
\ctikzset{bipoles/diode/height=0.3}
\ctikzset{bipoles/diode/width=.3}
\ctikzset{tripoles/thyristor/height=1}
\ctikzset{tripoles/thyristor/width=1}
\ctikzset{bipoles/vsourceam/height/.initial=.7}
\ctikzset{bipoles/vsourceam/width/.initial=.7}
\tikzstyle{every node}=[font=\small]
\tikzstyle{every path}=[line width=0.8pt,line cap=round,line join=round]

\usepackage{amssymb}
\usepackage{color}
\usepackage{textcomp}
\usepackage{comment}
\usepackage{xcolor}

\begin{document}

\title{A Neural-Network-Based Model Predictive Control of Three-Phase Inverter With an Output $LC$ Filter}

\author{\IEEEauthorblockN{Ihab S. Mohamed\IEEEauthorrefmark{1}, 
Stefano Rovetta\IEEEauthorrefmark{2}, Ton Duc Do \IEEEauthorrefmark{3}, Tomislav Dragi\u{c}evi\'{c} \IEEEauthorrefmark{4}, and Ahmed A. Zaki Diab \IEEEauthorrefmark{5} 
}\\
\IEEEauthorblockA{\IEEEauthorrefmark{1} INRIA Sophia Antipolis - M\'editerran\'ee, University C\^ote d'Azur, France (e-mail: ihab.mohamed@inria.fr)}\\
\IEEEauthorblockA{\IEEEauthorrefmark{2} Department of Informatics, Bioengineering, Robotics and Systems Engineering, University of Genoa, Italy (e-mail: stefano.rovetta@unige.it)}\\
\IEEEauthorblockA{\IEEEauthorrefmark{3} Department of Robotics and Mechatronics, School of Science and Technology (SST), Nazarbayev University, Astana Z05H0P9, Republic of Kazakhstan (e-mail: doduc.ton@nu.edu.kz)}\\
\IEEEauthorblockA{\IEEEauthorrefmark{4} Department of Energy Technology, Aalborg University, Denmark (e-mail: tdr@et.aau.dk)}\\
\IEEEauthorblockA{\IEEEauthorrefmark{5} Electrical Engineering Department, Faculty of Engineering, Minia University, Egypt (e-mail: a.diab@mu.edu.eg)}
}



\maketitle

\begin{abstract}
Model predictive control (MPC) has become one of the well-established modern control methods for three-phase inverters with an output $LC$ filter, where a high-quality voltage with low total harmonic distortion (THD) is needed. Although it is an intuitive controller, easy to understand and implement, it has the significant disadvantage of requiring a large number of online calculations for solving the optimization problem.
On the other hand, the application of model-free approaches such as those based on artificial neural networks approaches is currently growing rapidly in the area of power electronics and drives. This paper presents a new control scheme for a two-level converter based on combining MPC and feed-forward ANN, with the aim of getting lower THD and improving the steady and dynamic performance of the system for different types of loads. First, MPC is used, as an expert, in the training phase to generate data required for training the proposed neural network. Then, once the neural network is fine-tuned, it can be successfully used online for voltage tracking purpose, without the need of using MPC. The proposed ANN-based control strategy is validated through simulation, using MATLAB/Simulink tools, taking into account different loads conditions. Moreover, the performance of the ANN-based controller is evaluated, on several samples of linear and non-linear loads under various operating conditions, and compared to that of MPC, demonstrating the excellent steady-state and dynamic performance of the proposed ANN-based control strategy.

\end{abstract}


\section{Introduction}\label{Introduction}

\begin{figure*}[!ht]
\renewcommand{\figurename}{Fig.}
\begin{center}
\includegraphics[width=0.9\textwidth]{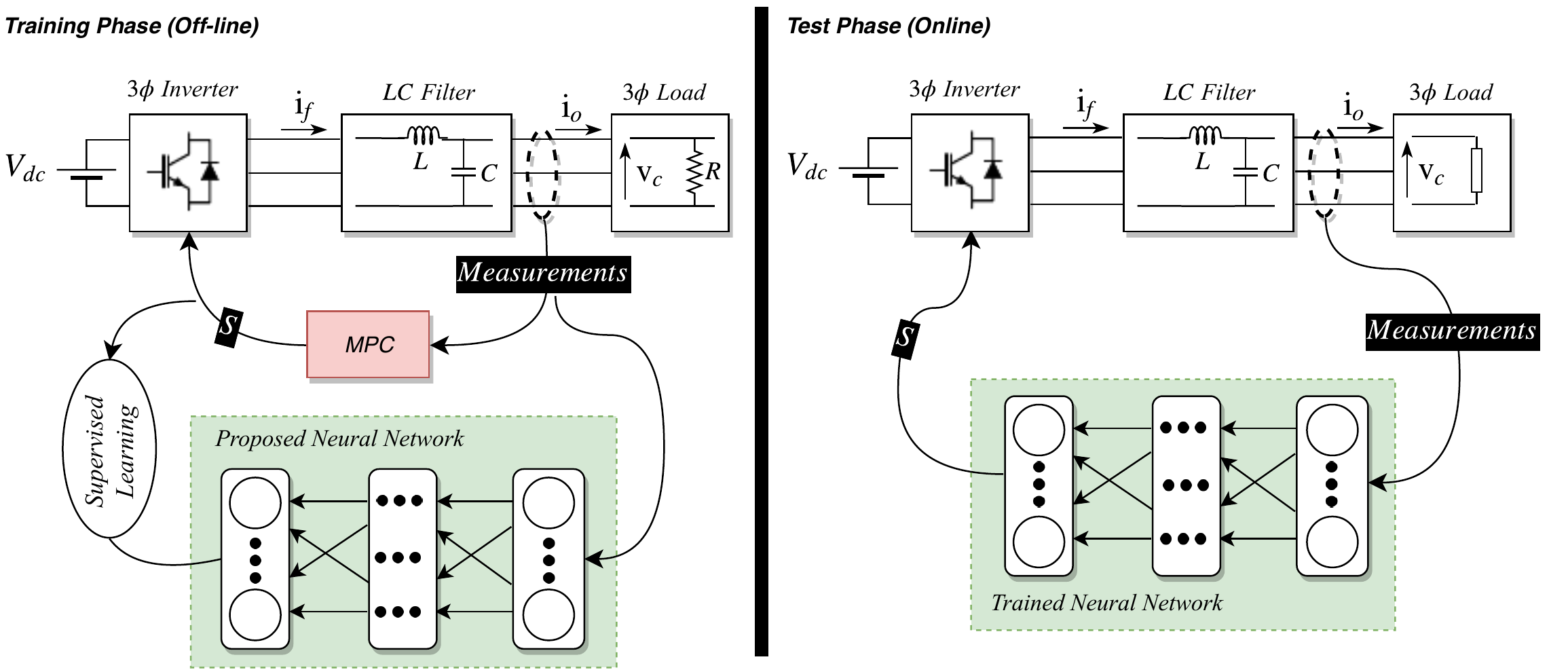}
\caption{An overview of the proposed control strategy: the training phase combines between using MPC for predicting the output voltage of the inverter and collecting data, under full-state observation, for training the neural network. In the test phase, the trained neural network is employed online to control the output voltage of the inverter instead of MPC, considering linear and non-linear loads.}
\label{fig:ANN-MPCdiagram}
\end{center}
\end{figure*}

\IEEEPARstart{T}{he} three-phase inverter is an extensively popular device, which is commonly used for transferring energy from a DC voltage source to an AC load. The control of three-phase inverters has received much attention in the last decades both in the scientific literature and in the industry-oriented research \cite{cortes2009model, habetler2002design}. In particular, for applications such as uninterruptible power supplies (UPSs), energy-storage systems, variable frequency drives, and distributed generation, the inverters are commonly used with an output $LC$ filter to provide a high-quality sinusoidal output voltage with low total harmonic distortion (THD) for various types of loads, especially for unbalanced or nonlinear loads \cite{kazmierkowski1998current, carrasco2006power, blaabjerg2006overview, gurrero2007uninterruptible}. However, the performance of the inverter is mainly dependent on the applied control technique. These controllers must cope with the load variations, the non-linearity of the system, and ensuring stability under any operating condition with a fast transient response \cite{hung1993variable}. 

In the literature, various types of classical and modern control schemes have been studied and proposed in order to improve the performance of the converters, such as non-linear methods (e.g., hysteresis voltage control (HVC)) \cite{mohamed2013classical}, linear methods (e.g., proportional-integral (PI)  controller with pulse-width modulation (PWM) and space vector modulation (SVM)) \cite{brod1985current, jung2004optimal, mohamed2013model, rojas2017new}, multi-loop feedback control \cite{loh2003comparative, loh2005analysis}, deadbeat control \cite{mohamed2007improved, lim2014robust, pichan2017deadbeat}, repetitive-based controllers \cite{escobar2007adaptive, jiang2012low}, linear quadratic controller (LQR) \cite{wu2005digital}, and sliding-mode control \cite{komurcugil2012rotating,  sabir2017robust}.

Most of these control schemes, in a way or another, are characterized by a number of limitations. For instance, the major drawback of non-linear methods (e.g., HVC), which require high switching frequency for effective operation, is having a variable switching frequency. This creates resonance problems which reduce the converter's efficiency \cite{cortes2008predictive, singh2018hil}. On the other hand, although the linear methods, which require carrier-based modulators, have the advantage of constant switching frequency, their dynamic response is weak comparing with HVC, because of the slow response of the modulator. However, both linear and nonlinear methods are extensively used for generating the switching signals of the inverter because of the simplicity of the controller implementation. Another example is deadbeat control which provides fast transient response, but is highly sensitive to model uncertainties, measurement noise, and parameter perturbations, in particular for high sampling rates. Other modern control approaches based on $H_{\infty}$ control theory \cite{willmann2007multiple} and $\mu$ synthesis \cite{lee2004robust} have been proposed, to handle the possible uncertainties in the system.

Model predictive control (MPC) has become one of the well-established modern control methods in power electronics, particularly for three-phase inverters with $LC$ filter according to \cite{cortes2008predictive, cortes2009model, nauman2016efficient, mohamed2016implementation, guo2019improved}. The key characteristic of MPC is to explicitly use the model of the system to predict the future behavior of the variables to be controlled, considering a certain time horizon. Afterwards, MPC selects the optimal control action (i.e., optimal switching signals) based on the minimization of a pre-defined cost function, which represents the desired behavior of the system \cite{camacho2007nonlinear, vazquez2017model, nguyen2017model}. With the aim of getting lower THD and improving steady and dynamic performance, many methods have been proposed in the literature \cite{vazquez2017fcs, guo2019improved}. For instance, the deployment of longer prediction horizons is presented in \cite{mohamed2015improved}. However, this results in a significant increase in computational cost. 
To mitigate and tackle this problem, an improvement of the finite-set FS-MPC strategy, using only a single step prediction horizon, is introduced in \cite{dragivcevic2017model}. This improvement is mainly based on defining a new cost function, which not only tracks the voltage reference but it also simultaneously tracks its derivative. 
While, in \cite{zheng2019current}, a current-sensorless FS-MPC scheme for LC-filtered voltage source inverters is proposed, in order to reduce the number of sensors in typical FS-MPC, offering a comparable performance with the typical FS-MPC scheme.

The main features of MPC can be summarized as:
(i) an intuitive controller easy to understand and implement, with a fast dynamic response; 
(ii) no need either for PWM blocks or modulation stage;
(iii) the simple inclusion of system constraints and nonlinearities, and multivariable cases;
(iv) the flexibility to include other system requirements.
On the other hand, a major drawback of MPC is that it requires the optimization
problem to be solved online, which involves a huge amount of real-time calculations. However, different solutions have been introduced in order to address this problem, as proposed in \cite{mariethoz2009explicit, kwak2014switching, nauman2016efficient}.

On the other hand, the application of data-driven methodologies (or model-free approaches, particularly artificial neural networks ANNs-based approaches) is currently growing rapidly in the area of power electronics and drives \cite{rashid2017power}. 
Broadly speaking, the use of neural networks for the control of dynamical systems was proposed in the early nineties \cite{ narendra1990identification, hunt1992neural, saint1991neural}. 
Multi-layer perceptrons were employed in various roles, including system identification and implementation of the control law. 
In particular, ANN-based controllers and estimators have been widely used in identification and control of power converters and motor drives	 \cite{karanayil2018artificial}. As an example, they can be used to estimate the rotor speed, rotor-flux, and torque of induction motors \cite{wishart1995identification, sun2013speed, lee2018performance}, in addition to the identification and estimation of the stator current of induction motor drives \cite{gadoue2013stator}. Several ANN-based methods have also been used in the control of power converters, as presented in \cite{bakhshai1996combined, pinto2000neural, karatepe2009artificial,akter2016modified}. Indeed, the ANN-based controllers have some advantages compared to other control methods such as: 
(i) their design does not require the mathematical model of the system to be controlled, considering the whole system as a black-box;
(ii) they can generally improve the performance of the system when they are properly tuned;
(iii) they are usually easier to be tuned as compared to conventional controllers;
(iv) they can be designed based on the data acquired from a real system or a plant in the absence of necessary expert knowledge. But, they require a large amount of training data. However, as the present work suggests, this is not a major drawback because data can be obtained using reliable simulation tools.

By taking advantage of the flexibility of MPC at training time, this paper proposes a feed-forward ANN-based controller for a three-phase inverter with output $LC$ filter for UPS applications. The goal is getting lower THD and good performance for different types of loads. The proposed controller undergoes two main steps: (i) we use MPC as an expert or a teacher for generating the data required for training off-line the proposed neural network using standard supervised learning, under full-state observation of the system;
(ii) once the off-line training is performed, the trained ANN can successfully control the output voltage of the inverter, without the need of using MPC at test time, as illustrated in Fig. \ref{fig:ANN-MPCdiagram}. We study a performance comparison between the proposed ANN-based approach and the conventional MPC, under various operating conditions. 
The main contributions of the work described in this
paper can be summarized as follows:
\begin{enumerate}
\item To the best of our knowledge, this is the first attempt to directly control a three-phase inverter with an output $LC$ filter using a feed-forward ANN based on MPC, instead of the more common model-based approaches as well as ANN classical control-based (such as Fuzzy Logic Controller FLC-, PID-, or PWM-based) approaches, or a combination of both \cite{pinto2000neural, lin1993power,sun2002analogue, boumaaraf2015three, wai2015design, fu2016control}. 
\item The proposed ANN-based approach generates directly the switching signals of the inverter, without the need for the mathematical model of the inverter and without a pre-defined cost function to be minimized at each sampling time $T_s$. This kind of approach is known as an end-to-end approach.
\item The proposed strategy exhibits very low computational cost compared to \cite{mohamed2015improved, dragivcevic2017model}, with much faster dynamic performance and significantly improved steady-state performance compared to conventional methods. 
\item An open repository of the dataset and codes is provided to the community for further research activities.\footnote{Web: \url{https://github.com/IhabMohamed/ANN-MPC}} 
\end{enumerate}

The rest of the paper is organized as follows. Section \ref{System Description and Modeling} deals with the mathematical model of the three-phase voltage-source inverter with $LC$ filter, whereas in Section \ref{MPC} the proposed predictive controller strategy is explained. The ANN-based control scheme proposed in this paper is described in Section \ref{ANN-MPC}. In Section \ref{SimulationAndResults}, simulation implementation and results are discussed for both proposed control schemes, then the conclusion is provided in Section \ref{CONCLUSION}.
\section{System Description and Modeling}\label{System Description and Modeling}
This section presents the mathematical interpretation of the converter system considered in this paper. The model of $LC$ filter is also described in details, and is then used by the predictive controller to predict the output voltage for all given input voltage vectors.
\subsection{System description via Clarke transformation}
\begin{figure}[h]
\renewcommand{\figurename}{Fig.}
\begin{center} 
\includegraphics[width=0.9\columnwidth]{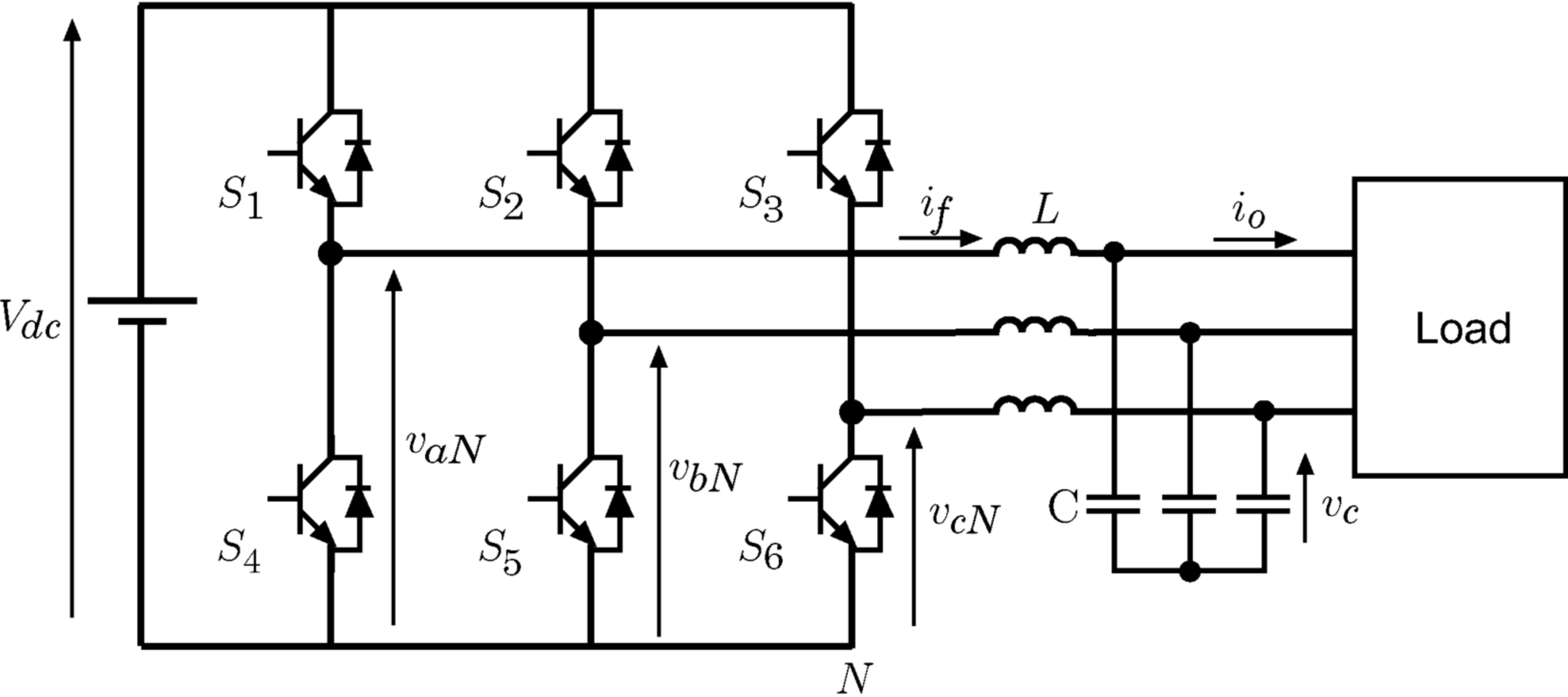}
\caption{Three-phase voltage-source inverter feeding an output $LC$ filter, which is directly connected to either linear or non-linear loads.}
\label{fig:system}
\end{center}
\end{figure}

The power circuit of the three-phase voltage-source inverter considered in this paper is depicted in Fig. \ref{fig:system}. 
In the present case, the load is assumed to be unknown, while the models of the converter and filter are given \cite{mohamed2013three}. Moreover, the two switches of each leg of the converter operate in a complementary mode, in order to avoid the occurrence of short-circuit conditions. Thus, the switching states of the converter can be represented by the three binary switching signals, $S_a$, $S_b$, and $S_c$, as follows: 
\begin{equation*}
S_a = \left\{ 
  \begin{array}{l l}
    1, & \quad \text{if $S_1$ ON and $S_4$ OFF}\\
    0, & \quad \text{if $S_1$ OFF and $S_4$ ON}\\
  \end{array} \right.
\end{equation*}
\begin{equation*}
S_b = \left\{ 
  \begin{array}{l l}
    1, & \quad \text{if $S_2$ ON and $S_5$ OFF}\\
    0, & \quad \text{if $S_2$ OFF and $S_5$ ON}\\
  \end{array} \right.
\end{equation*}
\begin{equation*}
S_c = \left\{ 
  \begin{array}{l l}
    1, & \quad \text{if $S_3$ ON and $S_6$ OFF}\\
    0, & \quad \text{if $S_3$ OFF and $S_6$ ON}\\
  \end{array} \right.
\end{equation*}

These switching states can be expressed in vectorial form (i.e., in $\alpha\beta$ reference frame) by following transformation:
\begin{equation}
\begin{split}
\textsc{S} &= \frac{2}{3}(S_a+\textsf{a} S_b +\textsf{a}^2 S_c) \equiv \textsc{S}_\alpha + j \textsc{S}_\beta, \\
\underbrace{\vphantom{
	\begin{bmatrix}
 		S_a\\
 		S_b\\
 		S_c 
 	\end{bmatrix}}	
\begin{bmatrix}
 		\textsc{S}_\alpha\\
 		\textsc{S}_\beta
	\end{bmatrix}}_{\textsc{S}} 
	&= \underbrace{
	\vphantom{
	\begin{bmatrix}
	  S_a\\	
	  S_b\\ 
	  S_c 
	  \end{bmatrix}}  
	\frac{2}{3}
	\begin{bmatrix}
 		1 & -1/2 & -1/2\\
 		0 & \sqrt{3}/2 & -\sqrt{3}/2
	\end{bmatrix}}_{=: T_{c}\;(\text{Clarke transformation})}
	\underbrace{\begin{bmatrix}
 		S_a\\
 		S_b\\
 		S_c 
 	\end{bmatrix}}_{S_{abc}},
\end{split}
\label{eq:sw}
\end{equation}
where $\textsf{a}=e^{j(2\pi/3)}$. The switching devices are assumed to be ideal switches, therefore the process of switching-ON/-OFF is not taken into consideration \cite{mohamed2016implementation}.

The possible output-voltage space vectors generated by the inverter can be obtained by
\begin{equation}
\text{v}_i=\frac{2}{3}(\boldsymbol{v}_{aN}+\textsf{a} \boldsymbol{v}_{bN}+\textsf{a}^2 \boldsymbol{v}_{cN})
\end{equation}
where $\boldsymbol{v}_{aN}$, $\boldsymbol{v}_{bN}$, and $\boldsymbol{v}_{cN}$ represent the phase-to-neutral, $N$, voltages of the inverter. On the other hand, we can define the voltage vector $\text{v}_i$ in terms of the switching state vector $\textsc{S}$ and the dc-link voltage $V_{dc}$ by 
\begin{equation}
\text{v}_i= V_{dc}\textsc{S}.
\label{eq:vi} 
\end{equation}

\begin{figure}[h]
\renewcommand{\figurename}{Fig.}
\centering
\includegraphics[scale=0.6]{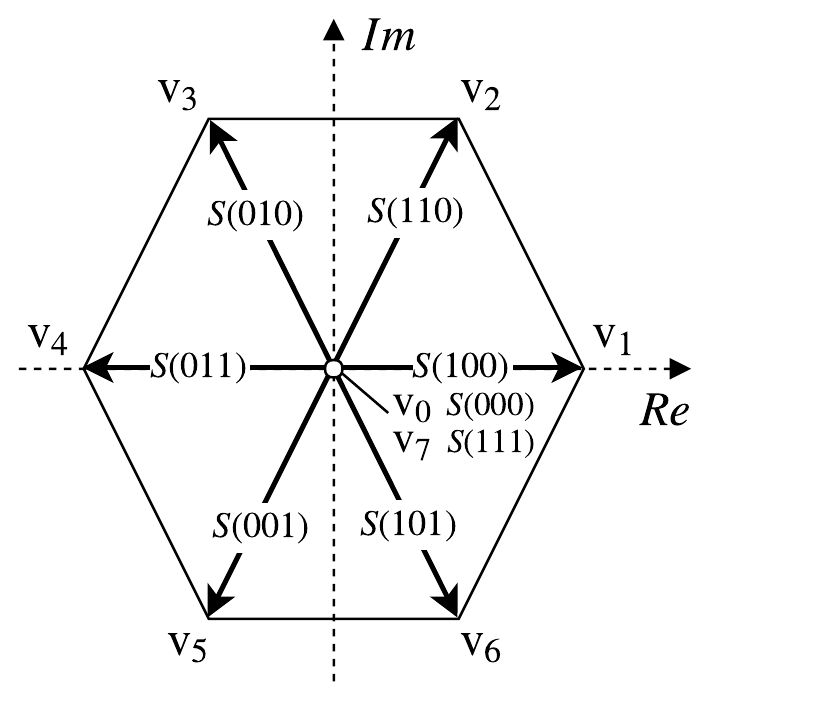}
\caption{Eight possible combinations of the switching signals, and their corresponding voltage vectors generated by the inverter in the complex $\alpha \beta$ frame.}
\label{fig:VoltVectors}
\end{figure}

Fig. \ref{fig:VoltVectors} illustrates the eight switching states and, consequently, the eight voltage vectors generated by the inverter using (\ref{eq:sw}) and (\ref{eq:vi}), considering all the possible combinations of the switching signals $S_a$, $S_b$, and $S_c$. It is noteworthy that only seven different voltage vectors are considered as possible outputs, since $\text{v}_0 = \text{v}_7$.

Similarly, as in (\ref{eq:sw}), the filter current $\text{i}_f$, the output
voltage  $\text{v}_c$, and the output current $\text{i}_o$ can be expressed in vectorial form as
\begin{equation}
\text{i}_f=\frac{2}{3}(\boldsymbol{i}_{fa}+\textsf{a} \boldsymbol{i}_{fb}+\textsf{a}^2 \boldsymbol{i}_{fc}) \equiv \text{i}_{f\alpha} + j \,\text{i}_{f\beta},
\end{equation}
\begin{equation}
\text{v}_c=\frac{2}{3}(\boldsymbol{v}_{ca}+\textsf{a} \boldsymbol{v}_{cb}+\textsf{a}^2 \boldsymbol{v}_{cc}) \equiv \text{v}_{c\alpha} + j \,\text{v}_{c\beta},
\end{equation}
\begin{equation}
\text{i}_o=\frac{2}{3}(\boldsymbol{i}_{oa}+\textsf{a} \boldsymbol{i}_{ob}+\textsf{a}^2 \boldsymbol{i}_{oc}) \equiv \text{i}_{o\alpha} + j \,\text{i}_{o\beta}.
\end{equation}

\subsection{$LC$ filter modeling}
The model of $LC$ filter can be described by two equations: the former describes the inductance dynamics, whereas the latter describes the capacitor dynamics \cite{cortes2009model}. These two equations can be written as a continuous-time state-space system as
\begin{equation*}
\frac{dx}{dt} = Ax + B \text{v}_i + B_{q} \text{i}_o,
\end{equation*}
\begin{equation}
\frac{d}{dt}
\underbrace{\begin{bmatrix}
\text{i}_f\\
\text{v}_c
\end{bmatrix}}_{x}
=
\underbrace{\begin{bmatrix}
  0 & -\frac{1}{L}\\
 \frac{1}{C} & 0 
\end{bmatrix}}_{A}
\underbrace{\begin{bmatrix}
\text{i}_f\\
\text{v}_c
\end{bmatrix}}_{x} 
+ 
\underbrace{\begin{bmatrix}
  \frac{1}{L}\\
 0 
\end{bmatrix}}_{B}
\text{v}_i
+ 
\underbrace{\begin{bmatrix}
  0\\
-\frac{1}{C} 
\end{bmatrix}}_{B_q}
\text{i}_o,
\label{eq:ContModel}
\end{equation}
where $L$ and $C$ are the filter inductance and the filter capacitance, respectively. The output voltage $\text{v}_c$ and the filter current $\text{i}_f$ can be measured, whilst the voltage vector $\text{v}_i$ can be calculated using (\ref{eq:vi}). The output current $\text{i}_o$ is considered as a disturbance due to its dependence on an unknown load, whereas the value of $V_{dc}$ is assumed to be fixed and known. The output voltage $\text{v}_c$ is considered as the output of the system, which can be written as a state equation as \(\text{v}_c =
\begin{bmatrix}   0 & 1 \end{bmatrix} x \).

Then, using (\ref{eq:ContModel}), the discrete-time state-space model of the filter can be obtained for a sampling time $T_s$ as
\begin{equation*}
x(k+1)=A_q x(k) + B_q \text{v}_i(k)+B_{dq} \text{i}_o(k),
\end{equation*}
\begin{equation}
\begin{aligned}
\underbrace{\vphantom{\int\limits^{T_s}_{0} e^{A\tau}B d\tau}
\begin{bmatrix}
\text{i}_f(k+1)\\
\text{v}_c(k+1)
\end{bmatrix}}_{x(k+1)}
&=
\underbrace{\vphantom{\int\limits^{T_s}_{0} e^{A\tau}B d\tau}
e^{AT_s}}_{A_q}
\underbrace{\vphantom{\int\limits^{T_s}_{0} e^{A\tau}B d\tau}
\begin{bmatrix}
\text{i}_f(k)\\
\text{v}_c(k)
\end{bmatrix}}_{x(k)} 
+ 
\underbrace{\int\limits^{T_s}_{0} e^{A\tau}B d\tau}_{B_q}
\text{v}_i(k)\\
&+
\underbrace{\int\limits^{T_s}_{0} e^{A\tau}B_d d\tau}_{B_{dq}}
\text{i}_o(k). 
\label{eq:DisModel}
\end{aligned}
\end{equation}

This model is used by the predictive controller (i.e., MPC) to predict the output voltage $\text{v}_c$ for all given input voltage vectors $\text{v}_i$. Then, for predicting the output voltage $\text{v}_c$ using (\ref{eq:DisModel}), we need the output current $\text{i}_o$ which can be estimated using (\ref{eq:io}), assuming that $\text{i}_o(k-1)=\text{i}_o(k)$ for sufficiently small sampling times $T_s$ as  proposed in \cite{cortes2009model, mohamed2015improved}.    
\begin{equation}
\text{i}_o(k-1)\cong\text{i}_o(k) = \text{i}_f(k-1)-\frac{C}{T_s}\Bigl(\text{v}_c(k)-\text{v}_c(k-1)\Bigr)
\label{eq:io}
\end{equation}

\section{Model Predictive Control for Neural Network}\label{MPC}
In this section we employ the model predictive control (MPC) proposed in \cite{mohamed2013three,vazquez2017model}, which provides the state-of-art of output-voltage control of three-phase inverter for UPS applications, for two purposes: 
(i) to generate the data required for the off-line training of the proposed neural network, and (ii) to compare its performance with the proposed ANN-based controller under linear and non-linear load conditions.

\subsection{Proposed Predictive Controller Strategy}
In the proposed control strategy, we assume that the inverter generates only a finite number of possible switching states and their corresponding output-voltage vectors, making it possible to solve the optimization problem of the predictive controller online \cite{cortes2009model}. MPC exploits the discrete-time model of the inverter to predict the future behavior of the variables to be controlled, for each switching state. Thereafter, the optimum switching state is selected, based on the minimization of a pre-defined cost function, and directly fed to the power switches of the converter in each sampling interval $T_s$, without the need for a modulation stage. We choose the cost function to be minimize  so as to achieve the lowest error between the predicted output voltage and the reference voltage. We express the cost function $J$, which defines the desired behavior of the system, in orthogonal coordinates by
\begin{equation}
J = 
\Bigl(\boldsymbol{v}^{*}_{c\alpha}-\boldsymbol{v}_{c\alpha}(k+1)\Bigr)^2
+
\Bigl(\boldsymbol{v}^{*}_{c\beta}-\boldsymbol{v}_{c\beta}(k+1)\Bigr)^2
\label{eq:costFun}
\end{equation}
where
$\boldsymbol{v}^{*}_{c \alpha}$
 and 
$\boldsymbol{v}^{*}_{c \beta}$
are the real and imaginary parts of the output-voltage reference vector $\text{v}^{*}_c$, while 
$\boldsymbol{v}_{c\alpha}$ and $\boldsymbol{v}_{c\beta}$ are the real and imaginary parts of the predicted output-voltage vector $\text{v}_c(k+1)$.

\begin{figure}[!h]
\renewcommand{\figurename}{Fig.}
\begin{center} 
\includegraphics[scale=0.815]{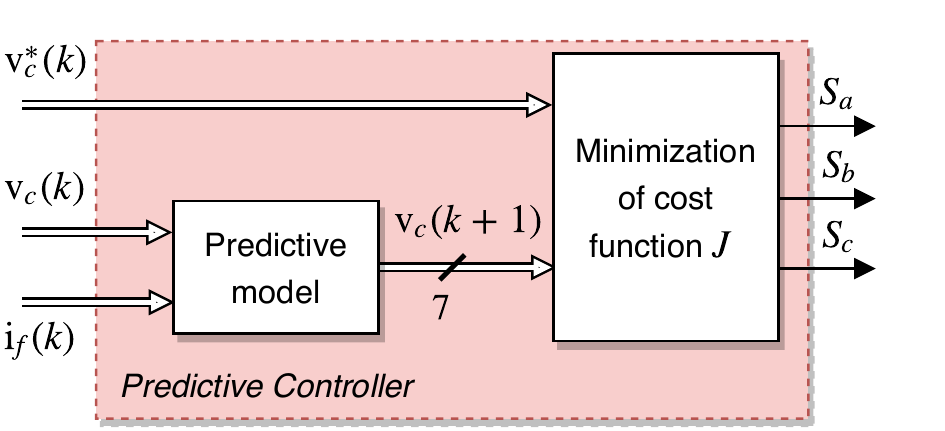}
\caption{Schematic diagram of the MPC scheme for a three-phase inverter with an output $LC$ filter. The controller takes the measured variables $\text{i}_f, \text{v}_c,$ and $\text{v}^{*}_c$ as inputs, while the switching signals $S_a$, $S_b$, and $S_c$ constitute the outputs.}
\label{fig:MPCdiagram}
\end{center}
\end{figure}

The block diagram of MPC, considering only one-step prediction horizon, for a three-phase inverter with output $LC$ filter is shown in Fig. \ref{fig:MPCdiagram}. The control cycle of the predictive controller at sampling instant $k$ is described as a pseudo code in Algorithm \ref{alg:MPCAlg.} with more detail. Line $1$ of the code declares the control function, where the switching signals $S_a$, $S_b$, and $S_c$ are the outputs, while the inputs are the measured variables of the filter current $\text{i}_f(k)$, the output voltage $\text{v}_c(k)$, and the reference voltage $\text{v}^{*}_c(k)$ at sampling time $k$, all expressed in $\alpha \beta$ coordinates. 
The two variables, $\text{i}_f(k-1)$ and $\text{v}_c(k-1)$, are recalled from the previous sampling instant (lines $7$ to $9$), which are firstly initialized for $k = 1$ (lines $3$ to $6$).
These two variables are used to estimate the output current $\text{i}_o(k)$ given by (\ref{eq:io}) (line $10$), in order to obtain the possible predictions of $\text{v}_c(k+1)$ using (\ref{eq:DisModel}).
 
The optimization is performed between lines $12$ and $20$.  
The code sequentially selects one of the seven possible voltage vectors $\text{v}_i$ generated by the inverter based on (\ref{eq:vi}) (line $13$) and applies it, in order to obtain the output voltage prediction $\text{v}_c(k+1)$ at instant $k + 1$, as in line $14$. The cost function given by (\ref{eq:costFun}) is used to evaluate the error between the reference and the predicted output voltage at instant $k + 1$ for each voltage vector (line $15$). The code selects the \textit{optimal} value of the cost function $J_{opt}$, and the \textit{optimum} voltage vector $x_{opt}$ is then chosen (lines $16$ to $19$). Note that $J_{opt}$ is initialized with a very high value (line $11$). Finally, the switching states, $S_a$, $S_b$, and $S_c$, corresponding to the \textit{optimum} voltage vector are generated and applied at the next sampling instant (line  $22$), as illustrated in Fig. \ref{fig:VoltVectors}. 

\begin{algorithm}[th!]
\caption{Pseudo code of the MPC scheme \cite{vazquez2017model}}
\label{alg:MPCAlg.}
\begin{algorithmic}[1]
\Function{$[S_a, S_b, S_c]=MPC(\textbf{\textit{i}}_f(k),\textbf{\textsc{v}}_c(k),\textbf{\textsc{v}}^{*}_c(k))$}{}
\State $\text{Measure the first sampled values as } \text{i}_f(1),\text{v}_c(1),\text{v}^{*}_c(1)$;
\If{$k = 1$}
	\State Set $\text{i}_f(k-1) = \text{i}_f(0) = 0+j0$;
	\State Set $\text{v}_c(k-1) = \text{v}_c(0) = 0+j0$;
\EndIf		
\If{$k > 1$}
	\State Recall measured variables $\text{i}_f(k-1),\text{v}_c(k-1)$;
\EndIf
\State Estimate $\text{i}_o(k) = \text{i}_f(k-1)-\frac{C}{T_s}\Bigl(\text{v}_c(k)-\text{v}_c(k-1)\Bigr)$;
\State Set $J_{opt}=\infty$;
\For{$l=1$ to $7$}
\State Compute $\text{v}_i(l)= \textsc{S}(l)V_{dc}$;
\State Predict $\text{v}_c(k+1)$ at instant $k + 1$ using (\ref{eq:DisModel});
\State Evaluate $J = \Bigl( \text{v}^{*}_c(k) - \text{v}_c(k+1)\Bigr)^2$;  
	\If{$J(l) < J_{opt}$}
		\State Set $J_{opt} = J(l)$;
		\State Set $x_{opt} = l$;	
	\EndIf	
\EndFor
\State Set $S_{opt}=S(x_{opt})$;
\State \textbf{return} $[S_a, S_b, S_c] = [S_{opt}(1), S_{opt}(2), S_{opt}(3)]$; 
\EndFunction
\end{algorithmic}
\end{algorithm}

\subsection{Discussion}
We can observe that all the control approaches proposed in the literature, in a way or another, are model-based approaches, which require in general either diverse computational or approximative procedures for applying their solution. In this context, MPC, the widely used approach for three-phase inverters, relies on solving an optimization problem online, leading to a large number of online computations. In other words, the control signal of MPC is determined by minimizing a cost function online at each time instant. Recently  artificial neural networks have been used in conjunction with MPC, in order to provide a powerful and fast optimization as proposed in \cite{piche2000nonlinear, aakesson2006neural, wang2018deep, dragicevic2018weighting}. 

The alternative approach considered in the present work is to apply neural network-based function approximators, which can be trained off-line to represent the optimal control law. Such an approach is expected to avoid the drawbacks associated with MPC-based control approaches, does not require the mathematical model of the system to be controlled, does not evaluate a cost function online at each sampling time, and, therefore, does not rely on an optimization problem to be solved online. For this reason, this paper focuses on the control of a three-phase inverter with output $LC$ filter using a feed-forward ANN-based MPC, which has not been reported in the literature, where MPC is only used as a teacher for training the neural network.

\section{Implementation of ANN-Based Controller}\label{ANN-MPC}
In this section, some important concepts related to ANN including the structure of the proposed ANN-based controller as well as details on the training data will be covered.

\subsection{Proposed Neural Network Architecture}
\begin{figure*}
\renewcommand{\figurename}{Fig.}
\begin{center} 
\includegraphics[scale=0.98]{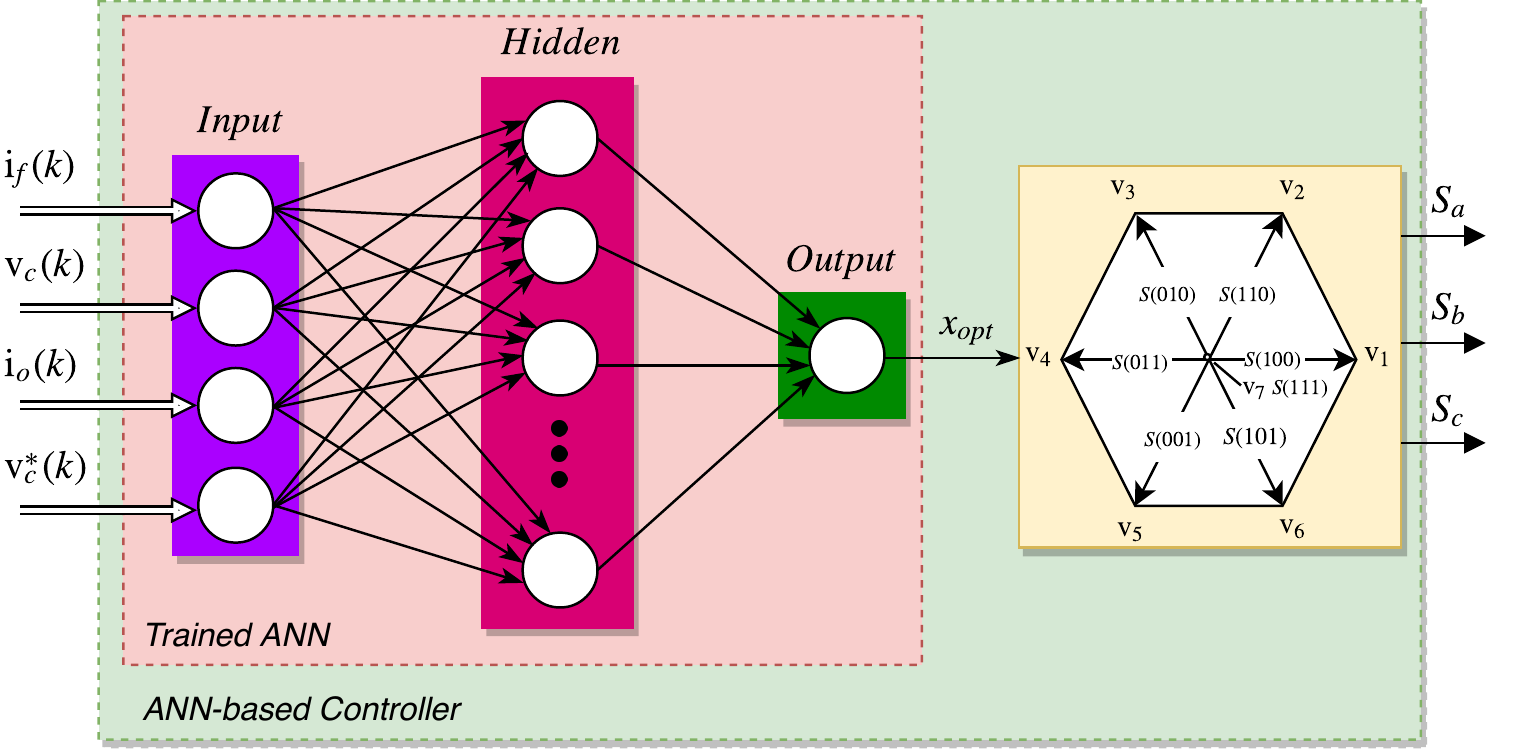}
\caption{Block diagram of the proposed ANN-based controller for a three-phase inverter with an output $LC$ filter. Each sampling instant, the trained ANN takes, as input, the measured variables $\text{i}_f, \text{v}_c, \text{i}_o$, and $\text{v}^{*}_c$, whereas it explicitly generates the \textit{optimum} voltage vector $x_{opt}$. Afterwards, the corresponding switching states $S_a$, $S_b$, and $S_c$ are directly given to the power switches of the converter.}
\label{fig:ANN-Controller}
\end{center}
\end{figure*}

Machine learning, and in particular artificial neural networks, is one key technology in modern control systems. An artificial neural network (ANN) is an extremely flexible computational model that can be optimized to learn input-to-output mappings based on historical data. 
An ANN is composed of a number of simple computing elements linked by weighted connections. Feed-forward networks do not contain loops, so they are organized in layers and can be used to implement input-to-output mappings that are memoryless, i.e., without dynamics. 
In its basic form, this model can be expressed as an iterative composition of input-output functions of the form
\begin{equation}
f(\vec{x}) = h\Bigl(w_0+\sum_{i=1}^{M} w_i x_i\Bigr),
\end{equation}
where $h(x)$ is an activation function (usually it is a non-linear function such as \textit{logistic sigmoid} or \textit{hyperbolic tangent}, to ensure the universal approximation property \cite{hornik1991approximation}), $\vec{x}=\{x_1,x_2, \cdots, x_M\}$	is the input vector of the ANN with $M$ elements, $w_i$ are the weights for each input $x_i$, and $w_0$ is a bias or correction factor. 
In a feed-forward network, it is possible to distinguish one input layer, one output layer, and hidden layers that connect the input to the output.
The objective of the ANN training phase is to optimize some cost function by finding optimal values for the $w_i$ and $w_0$.

Although recent developments have focused on larger and larger scale problems (deep learning), improved techniques have also been proposed to improve the reliability of networks of smaller size. Toward the same goal, hardware suppliers have started to support reduced-precision floating-point \cite{FP16:2019} and integer \cite{Int16:2018} arithmetics, and offer small-scale, dedicated architectures\cite{Hu:2019}. The result is a sound and scalable technology.

In this work, a feed-forward neural network (fully connected multi-layer perceptron) of the ``shallow'' type, i.e., one hidden layer, was used to implement the control model. A grid search tuning procedure allowed the selection of a configuration with 15 units in the hidden layer, while the number of input and output units is constrained by the number of input and output variables, respectively. Training was done via the Scaled Conjugate Gradient (SCG) method \cite{Moller1993}, which exploits the good convergence properties of conjugate gradient optimization \cite{Fletcher2000} and has the computational advantage of not requiring a line search, nor any user-selected parameters.

\subsection{ANN Training Procedure}
The ANN takes as inputs the measured variables of the filter current $\text{i}_f$, the output voltage $\text{v}_c$, the output current $\text{i}_o$, and the reference voltage $\text{v}^{*}_c$ all expressed in $\alpha \beta$ coordinates. The real and imaginary parts of these variables are separately fed to the neural network, bringing the total number of input features to eight, i.e., $M =8$.
The output of the ANN is the \textit{optimum} voltage vector $x_{opt}$ to be applied at each sampling instant. The size of the output layer is an array with a length of $7$, which represents the indexes of the seven possible voltage vectors $\text{v}_i$ that inverter generates. The output is one-hot encoded, meaning that at each sampling instant only the index of the \textit{optimum} voltage vector will be active (i.e., having a value of one), while others will be equal to zero.

The training data, which have been collected by MPC, comprises $70$ experimental conditions, which are divided into $60$ cases for specific resistive loads (i.e., for only $R = 1, 3, 5, 7, 10, 15, 20, 25, 30,$ and $35$ $\Omega$), whereas only $10$ experiments represent the case where the inverter directly feeds a non-linear load (i.e., diode-bridge rectifier) with different values of $R_{NL}$ and $C_{NL}$. 
For each experimental condition, the simulation is run using MPC\footnote{Web: \url{https://github.com/IhabMohamed/MPC-3-Phase-Inverters}}, under various operating conditions such as simulation time (i.e, number of output voltage cycles), sampling time $T_s$, filter capacitor $C$, filter inductance $L$, DC-link voltage $V_{dc}$, and reference voltage $\text{v}^{*}_c$. Then, the input features of the neural network and their targets are stored for training. 

As a consequence, the total dataset consists of $217,510$ and $247,820$ instances for the cases where $60$ and $70$ experimental conditions are used, respectively. These dataset has been divided into two parts: $70\%$ randomly selected for training purposes, and $30\%$ for testing and validation. The overall accuracy of ANN for the $60$ training cases is $69.1\%$, while it has a $0.2\%$ increase for the $70$ training cases, considering $15$ hidden layers and the training function ``\textit{transcg}''. 
We observe that the validation and training error, as well as the error on the test set, are very similar when training stops, according to the ``\textit{early stopping}'' criterion used. This is an indication that the neural network may attain a good degree of generalization. For instance, for the $60$ training cases, the best validation performance is taken from epoch $747$ with the lowest validation error of $0.11108$. The training results are summarized in Table \ref{table:training}. 
Training was also attempted using the Bayesian regularization back-propagation method, achieving an accuracy of $93\%$. However, its performance at the test phase (on-line) was not satisfactory.
\begin{table}[!ht]
\caption{Training results of the proposed ANN based on 60 and 70 training cases, which have been collected by MPC}
\centering
\small\addtolength{\tabcolsep}{-4pt}
\begin{tabular}{l||c|c|c|} 
\hline 
 Tr. Cases & No. of Instances & Accuracy & Validation Error (epoch)\\
 \hline \hline  
 $60$    &  $217,510$ & $69.1\%$ & $0.11108$ (747)\\
 \hline
 $70$    & $247,820$ & $69.3\%$  & $0.11213$ (526)\\
\hline
\end{tabular}
\label{table:training}
\end{table}

For further detailed information about the training cases used for training the ANN-based controller, please refer to: 
\url{https://github.com/IhabMohamed/ANN-MPC}. 

\subsection{ANN-Based Controller}\label{ANN}

As previously mentioned, the ANN-based controller is trained off-line from samples collected via MPC, as shown in Fig. \ref{fig:ANN-MPCdiagram}. After fine-tuning the ANN, the trained ANN can be used instead of MPC to control the system presented in Fig. \ref{fig:system}.

Fig. \ref{fig:ANN-Controller} depicts the proposed block diagram of the ANN-based controller for a three-phase inverter with output $LC$ filter, in order to generate a high-quality sinusoidal output voltage with low THD, considering different types of loads.

The control strategy of the proposed ANN-based controller at sampling time $k$ can be described as follows:
\begin{enumerate}
\item measure the value of the filter current $\text{i}_f(k)$, the output voltage $\text{v}_c(k)$, and the output current $\text{i}_o(k)$ at sampling time $k$. Note that, the output current $\text{i}_o(k)$ is considered to be a measurable value, without estimation based on (\ref{eq:io}) or using the observer as in \cite{cortes2009model}; 
\item then, these measured values in addition to the reference voltage $\text{v}^{*}_c(k)$ are used by the trained ANN in order to explicitly generate the \textit{optimum} voltage vector $x_{opt}$ to be applied at instant $k + 1$; 
\item finally, the switching states, $S_a$, $S_b$, and $S_c$, corresponding to the optimum voltage vector $x_{opt}$ are applied and directly given to the power switches of the converter each sampling interval $T_s$. 
\end{enumerate}
\section{Simulation Implementation and Results}\label{SimulationAndResults}
This section provides a comprehensive study and evaluation of the two proposed control strategies, taking into account different loads under various operating conditions. 

\subsection{Simulation Setup}
To verify the proposed ANN-based control strategy and compare its performance with the conventional MPC,
we used MATLAB (R2018a)/Simulink software components to implement the Simulink model and the simulations of the system shown in Fig. \ref{fig:system}.
We acquired the training samples, off-line training, and online voltage tracking purpose using the proposed ANN approach via a PC equipped with an Intel\textsuperscript{\textregistered} Core i5-4210U $1.70$ $GHz$ CPU, $6$ GB of RAM, and an Nvidia Geforce\textsuperscript{\textregistered} GPU, and running Ubuntu 16.04 $64$ bit.

\subsection{Simulation Results}
The simulation of the three-phase inverter system shown in Fig. \ref{fig:system} was carried out, considering linear (i.e., resistive) and non-linear loads, in order to evaluate the behavior of the proposed ANN-based control strategy and compare its performance with that of MPC proposed in Section \ref{MPC}. In particular, we studied and evaluated the steady and dynamic performance of both control strategies, taking into account different loads conditions. The parameters of the system are listed in Table \ref{table:SysParameters}.

\begin{table}[!ht]
\caption{Parameters of the converter system}
\centering
\begin{tabular}{l c} 
\hline
 Parameter & Value \\
 \hline  
 DC-link voltage $V_{dc}$ & \SI{500}{\;[\volt]}\\
 Filter capacitor $C$ & \SI{40}{\;[\micro\farad]}\\
 Filter inductance $L$ & \SI{2}{\;[\milli\henry]}\\
 Sampling time $T_s$ & \SI{30}{\;[\micro\second]}\\
\hline
\end{tabular}
\label{table:SysParameters}
\end{table}

\begin{figure}[h!]
\renewcommand{\figurename}{Fig.}
\begin{center} 
\includegraphics[scale=0.98]{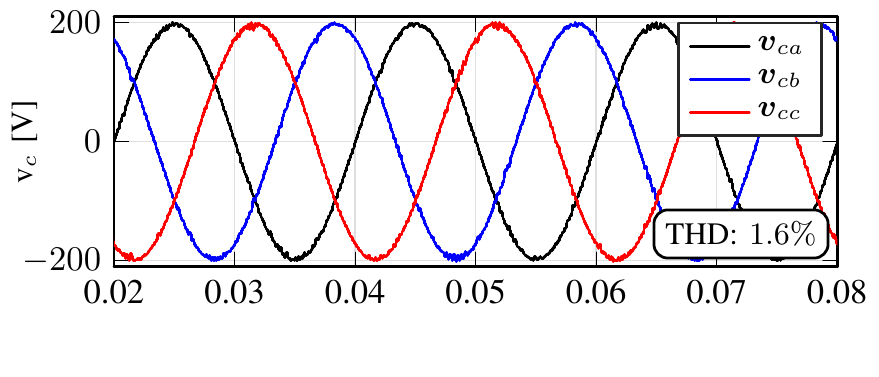}\\
\vspace{-0.5cm}
\includegraphics[scale=0.98]{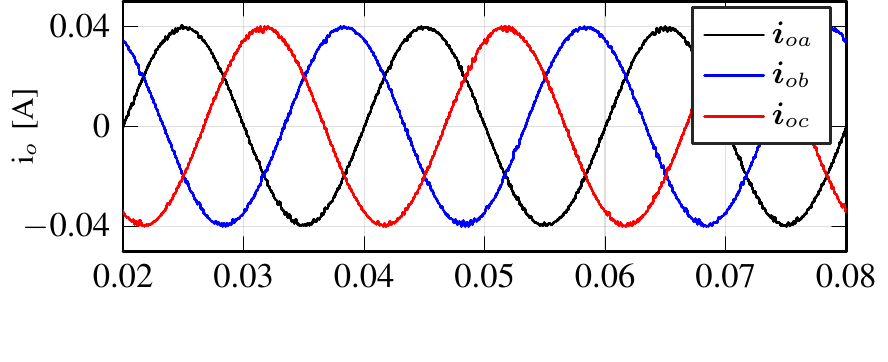}\\
\vspace{-0.5cm}
\includegraphics[scale=0.98]{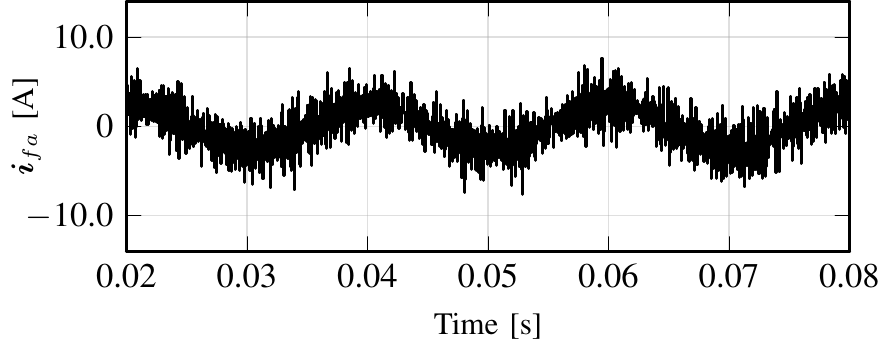}
\caption{Simulation results of ANN-based controller: output voltages, output currents, and filter current in steady-state for a resistive load of \SI{5}{\kilo\ohm}.}
\label{fig:R5k-ANN}
\end{center}
\end{figure}

\begin{figure}[h!]
\renewcommand{\figurename}{Fig.}
\begin{center} 
\includegraphics[scale=0.98]{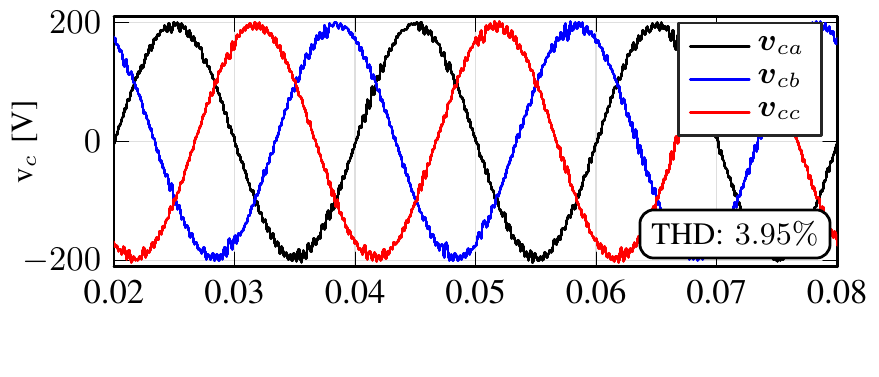}\\
\vspace{-0.5cm}
\includegraphics[scale=0.98]{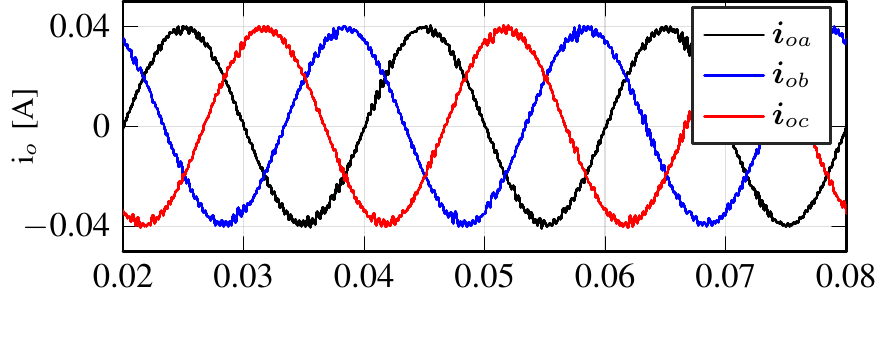}\\
\vspace{-0.5cm}
\includegraphics[scale=0.98]{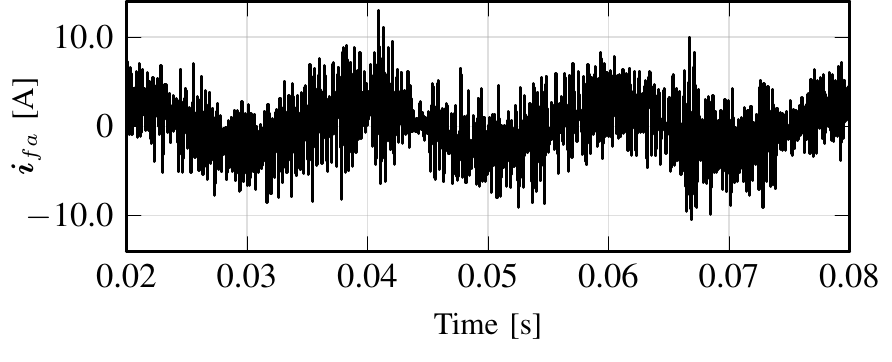}
\caption{Simulation results of MPC: output voltages, output currents, and filter current in steady-state for a resistive load of \SI{5}{\kilo\ohm}.}
\label{fig:R5k-MPC}
\end{center}
\end{figure}

The behavior of the ANN-based controller in steady-state operation for a resistive load of \SI{5}{\kilo\ohm} shown in Fig. \ref{fig:R5k-ANN}, while the behavior of the predictive controller for the same resistive load is shown in Fig. \ref{fig:R5k-MPC}. The amplitude and the fundamental frequency of reference voltage $\text{v}^{*}_c$ are set to \SI{200}{\volt} and \SI{50}{\Hz}, respectively. It can be seen in the figures that the output voltages $\text{v}_c$ for the proposed control strategies are sinusoidal with low distortion, particularly for the ANN-based approach which has a THD of only $1.6\%$ compared to $3.95\%$ for MPC. Moreover, we observe that, due to the resistive load, the output current $\text{i}_o$ is proportional to the output voltage, whilst the filter current $\text{i}_o$ measured at the output of the converter shows high-frequency harmonics, especially in the case of MPC, which are attenuated by the $LC$ filter. 

\begin{figure}[!ht]
\renewcommand{\figurename}{Fig.}
\begin{center} 
\includegraphics[scale=0.98]{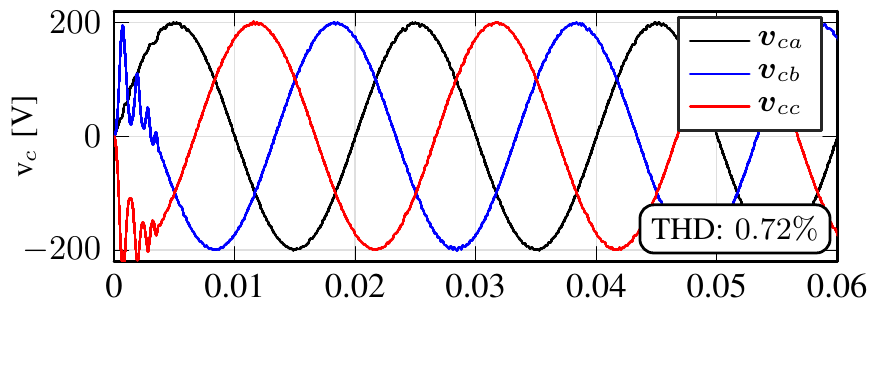}\\
\vspace{-0.5cm}
\includegraphics[scale=0.98]{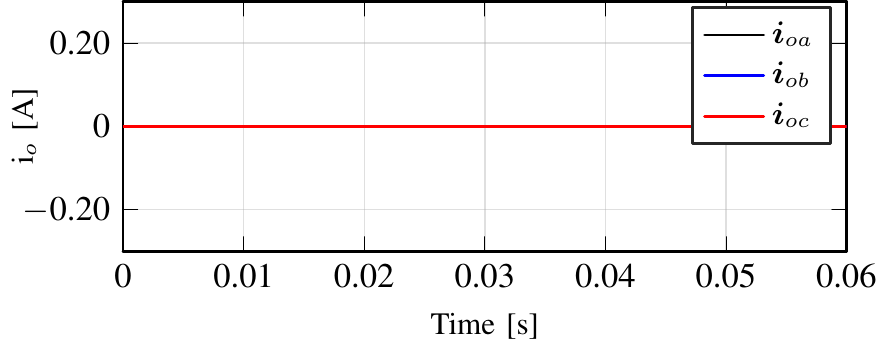}
\caption{Simulation results: the dynamic response of the ANN-based controller for a no-load, where the filter capacitor $C=\SI{50}{\micro\farad}$, the filter inductance $L=\SI{3.5}{\milli\henry}$, and $ T_s=\SI{30}{\micro\second}$.}
\label{fig:NoLoad-ANN}
\end{center}
\end{figure}

\begin{figure}[!ht]
\renewcommand{\figurename}{Fig.}
\begin{center} 
\includegraphics[scale=0.98]{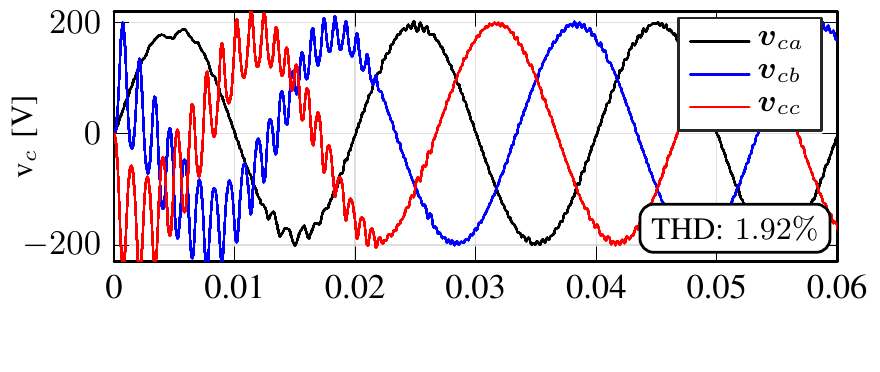}\\
\vspace{-0.5cm}
\includegraphics[scale=0.98]{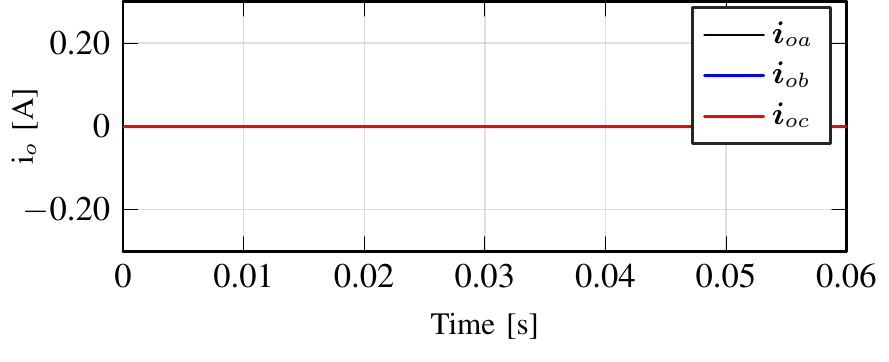}
\caption{Simulation results: the dynamic response of MPC for a no-load, where the filter capacitor $C=\SI{50}{\micro\farad}$, the filter inductance $L=\SI{3.5}{\milli\henry}$, and $T_s=\SI{30}{\micro\second}$.}
\label{fig:NoLoad-MPC}
\end{center}
\end{figure}

The transient response of both control strategies for no-load (i.e., open-circuit) is shown in Fig. \ref{fig:NoLoad-ANN} and Fig. \ref{fig:NoLoad-MPC}. Here, the filter capacitor $C$ and filter inductance $L$ are set to \SI{50}{\micro\farad} and \SI{3.5}{\milli\henry}, respectively, whilst the sampling time $T_s$ is kept constant at a value of \SI{30}{\micro\second}. It can be seen that the ANN-based controller permits a fast and safe transient response, demonstrating the excellent
dynamic performance of the proposed ANN-based control strategy. For MPC, the time elapsed in order to reach steady-state operation and to faithfully track its reference waveform is about \SI{20}{\milli\second} ($1$ \textit{cycle}), which is affected by the change in the load, as illustrated in Table \ref{table:Comparison}. On the other side, for the ANN-based controller, it is observed that it takes less than \SI{5}{\milli\second} for any load, in order to reach steady-state. Furthermore, the output voltage quality of ANN-based approach is improved significantly, with a THD of $0.72\%$ compared to $1.92\%$ for MPC.

\begin{figure}[!ht]
\renewcommand{\figurename}{Fig.}
\begin{center} 
\includegraphics[scale=0.13]{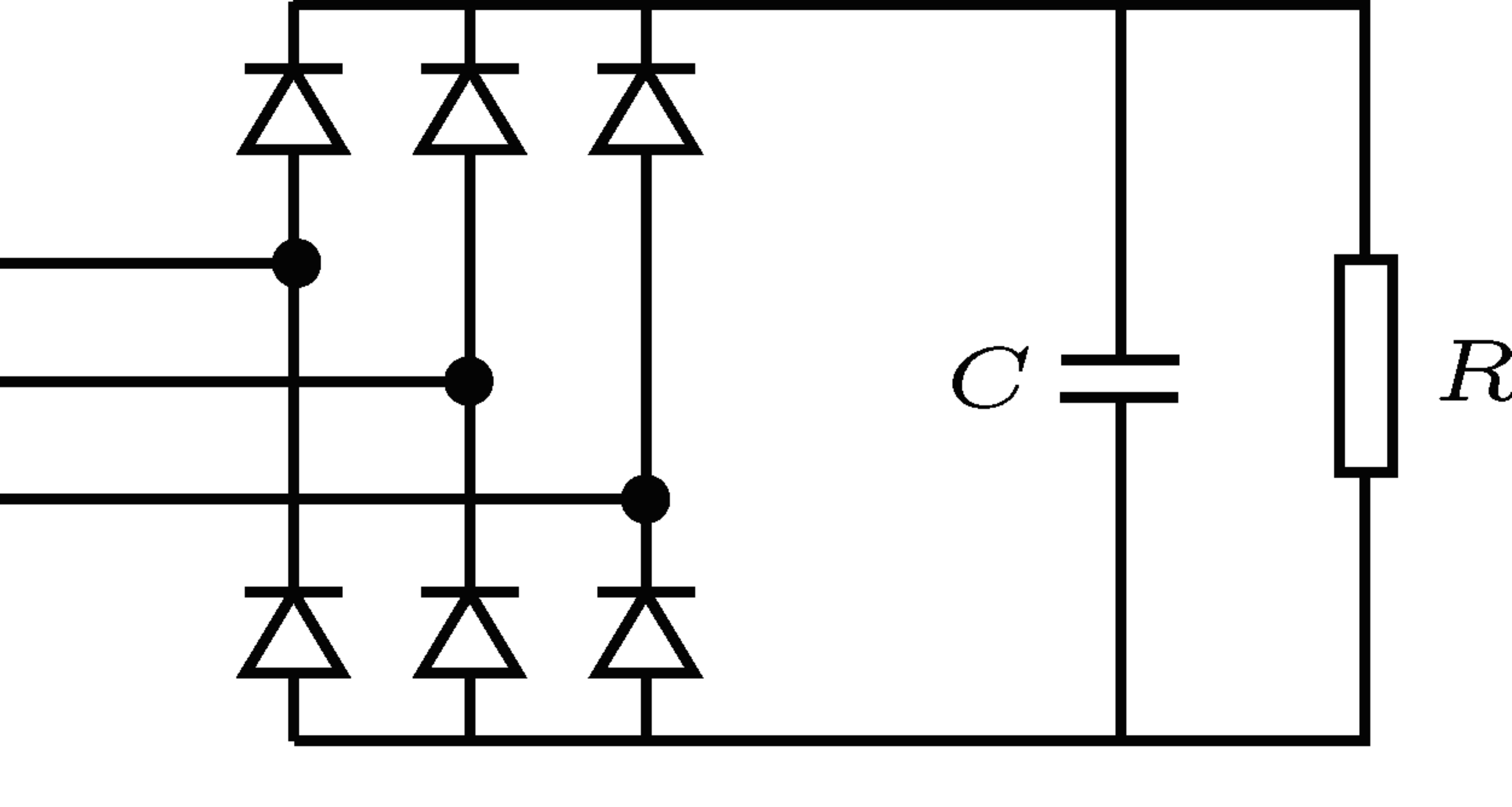}
\caption{Diode-bridge rectifier used as non-linear load, with values $C=\SI{300}{\micro\farad}$ and $R= \SI{60}{\ohm}$.}
\label{fig:Nonlinear}
\end{center}
\end{figure}

\begin{figure}[!ht]
\renewcommand{\figurename}{Fig.}
\begin{center} 
\includegraphics[scale=0.98]{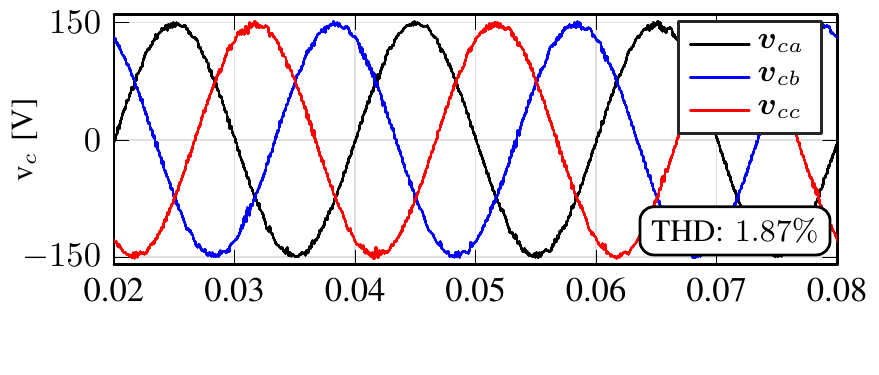}\\
\vspace{-0.5cm}
\includegraphics[scale=0.98]{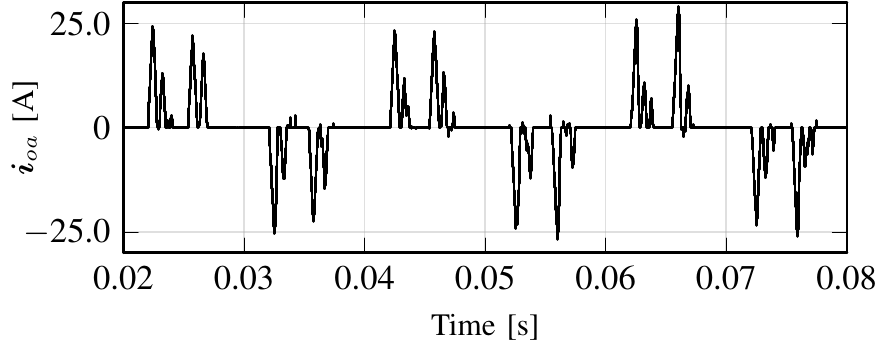}
\caption{Simulation results of ANN-based controller: output voltages and one-phase output current in steady-state for a diode-bridge rectifier and a reference amplitude of \SI{150}{\volt}.}
\label{fig:NonLinear-ANN}
\end{center}
\end{figure}

\begin{figure}[!ht]
\renewcommand{\figurename}{Fig.}
\begin{center} 
\includegraphics[scale=0.98]{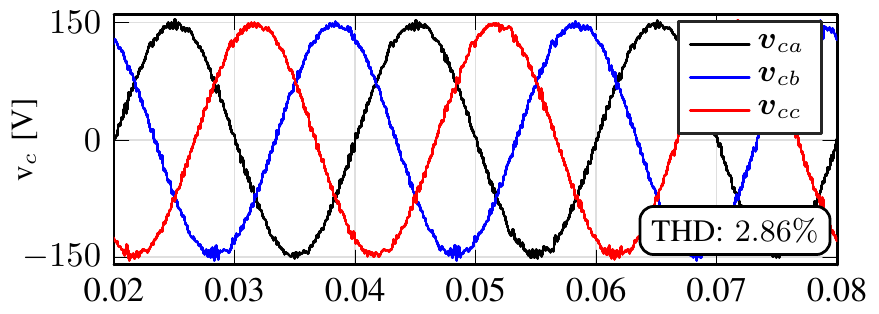}\\
\vspace{0.1cm}
\includegraphics[scale=0.98]{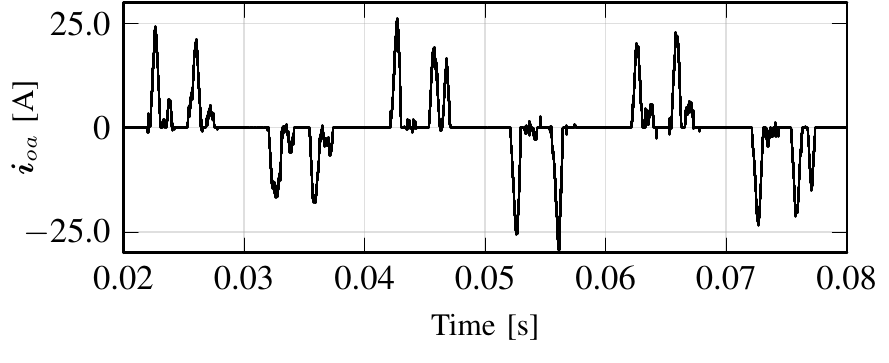}
\caption{Simulation results of MPC: output voltages and one-phase output current in steady-state for a diode-bridge rectifier and a reference amplitude of \SI{150}{\volt}.}
\label{fig:NonLinear-MPC}
\end{center}
\end{figure}

\begin{figure}[!ht]
\renewcommand{\figurename}{Fig.}
\begin{center} 
\includegraphics[scale=0.98]{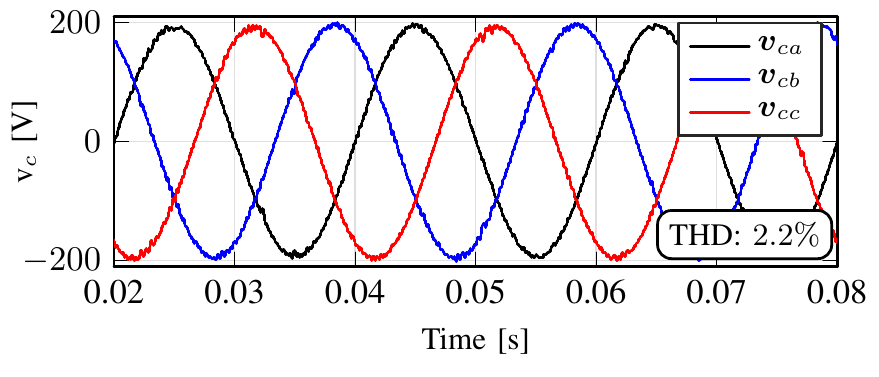}
\caption{Simulation results of ANN-based controller: output voltages in steady-state for an inductive load of \SI{0.01}{\henry} and a reference amplitude of \SI{200}{\volt}.}
\label{fig:Ind-ANN}
\end{center}
\end{figure}

\begin{figure}[!ht]
\renewcommand{\figurename}{Fig.}
\begin{center} 
\includegraphics[scale=0.98]{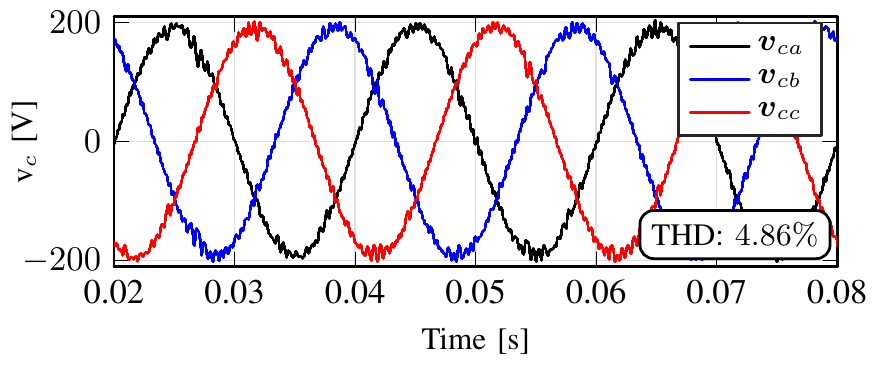}
\caption{Simulation results of MPC: output voltages in steady-state for an inductive load of \SI{0.01}{\henry} and a reference amplitude of \SI{200}{\volt}.}
\label{fig:Ind-MPC}
\end{center}
\end{figure}

As previously mentioned, the proposed ANN is trained off-line using a dataset which represents only different values of resistive load under different operating conditions. However, to verify the feasibility and effectiveness of the proposed ANN-based controller under realistic conditions, the behavior of the system is tested online considering non-linear loads, such as a diode-bridge rectifier as shown in Fig. \ref{fig:Nonlinear} and an inductive load. Fig. \ref{fig:NonLinear-ANN} and Fig. \ref{fig:NonLinear-MPC} show the behavior of the proposed control strategies for a diode-bridge rectifier, with values $C=\SI{300}{\micro\farad}$ and $R= \SI{60}{\ohm}$, while the behavior for an inductive load of \SI{0.01}{\henry} is shown in Fig. \ref{fig:Ind-ANN} and Fig. \ref{fig:Ind-MPC}, considering the same operating conditions presented in Table \ref{table:SysParameters} and different amplitudes of the reference output voltage. As can be seen in the figures, the output voltage generated by the ANN-based controller outperforms that obtained using MPC for non-linear loads, despite the highly distorted output currents due to feeding a non-linear load. For instance, for MPC, the total distortion in the output voltage for the inductive load was $4.86\%$, while it was $2.2\%$ for the ANN-based controller. The result of MPC can be improved by using either a smaller sampling time or a higher value of the filter capacitance \cite{mohamed2016implementation}. 
\begin{figure*}[!t]
\renewcommand{\figurename}{Fig.}
\begin{center} 
\includegraphics[scale=0.95]{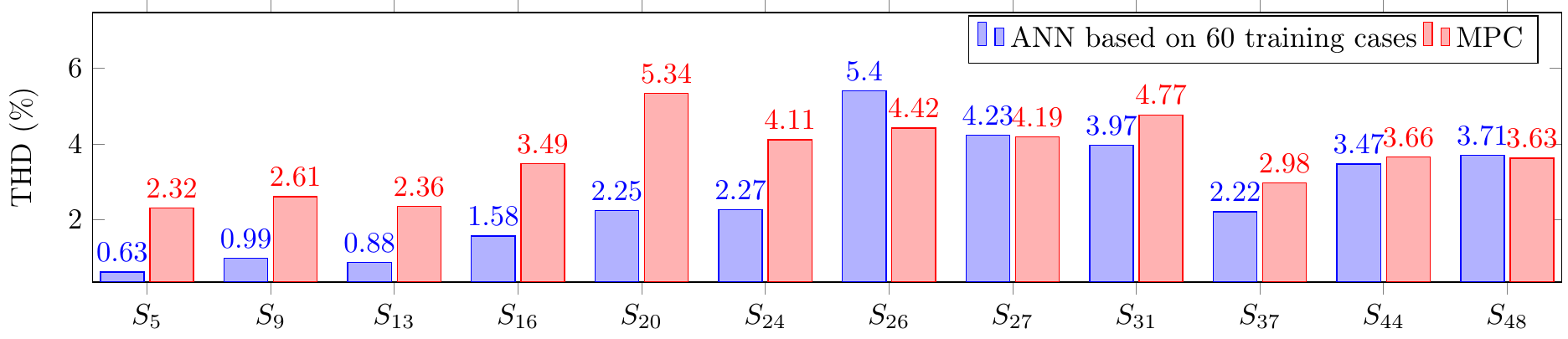}
\caption{Comparison of the THD of the output voltage obtained by the two proposed control strategies, for some cases given in Table \ref{table:Comparison}, under different operating conditions.}
\label{fig:perf}
\end{center}
\end{figure*}

In order to achieve a fair comparison and prove the superiority of the proposed ANN-based approach compared to MPC in both transient and steady-state response, Table \ref{table:Comparison} shows a comprehensive comparison of both the control strategies  for linear and non-linear loads, under various operating conditions such as sampling time $T_s$, filter capacitor $C$, filter inductance $L$, DC-link voltage $V_{dc}$, and reference voltage $\text{v}^{*}_c$. Fifty unseen cases, at training time, have been considered for testing the proposed approaches, including thirty cases for different values of a resistive load, whereas the rest was for a diode-bridge rectifier as a non-linear load. 
Moreover, the THD of the output voltage obtained by the proposed control strategies, for some cases given in Table \ref{table:Comparison}, is visualized in Fig. \ref{fig:perf}. As anticipated, the performance of the ANN-based approach, either based on sixty or seventy training cases, outperforms that of MPC, which can be noticed in lower THD and less settling time to reach steady-state (i.e., $t_{ss}$, as shown in the first ten samples (i.e., $S_1-S_{10}$)). It can be noticed that the performance of the ANN-based controller using only sixty training cases is similar to that based on seventy cases (see column $8$ and $9$ in Table \ref{table:Comparison}). 

However, for cases $S_{26}-S_{30}$, the output voltages obtained using MPC are better than that obtained using the ANN-based controller. Moreover, it can be seen in sample $S_{49}$ that the ANN-based approach failed to control the output voltage and track its reference waveform. As a consequence, the UPS does not work properly due to a higher distortion in the voltage. These results could be improved using either (i) a higher sampling frequency, or (ii) a higher value of the filter capacitance $C$, as illustrated in sample $S_{50}$ which represents an improvement of the result of sample $S_{49}$. An alternative solution to be considered to improve the performance of the controller is to increase the number of training instances, taking into account various values of $C$ and $T_s$. In addition, it is observed that having a  one-delay step in the input features of the neural network improves its performance to outperform that of MPC. For example, $(THD)_{ANN}$ of cases $S_{26}, S_{27}, S_{28}, S_{29}, S_{49}$ is decreased to be $3.72\%, 2.39\%, 4.08\%, 2.35\%, 3.86\%$, respectively.

In fact, it is not surprising that the performance of the proposed ANN-based controller outperforms that of MPC in both transient and steady-state response, even with unseen experimenta conditions (i.e., loads) at training time as tabulated in Table \ref{table:Comparison}. This happened for two reasons. First, the training data are sufficient to learn the mathematical model of the system to be controlled and its dynamics, as well as representing the optimal control law. Second, generating a sinusoidal output voltage can be considered as a repetitive task, where neural network can easily detect and learn repetitive sequences of actions. 

\begin{table*}[!ht]
\caption{A comparison between the two proposed control strategies for linear and non-linear loads under different operating conditions such as sampling time $T_s$, filter capacitor $C$, filter inductance $L$, DC-link voltage $V_{dc}$, and reference voltage $\text{v}^{*}_c$}
\centering
\small\addtolength{\tabcolsep}{-2.5pt}
\begin{tabular}{|c||c|c|c|c|c|c|c||c|c|c|}
\hline
\multicolumn{8}{|l||}{\multirow{2}{*}{\textcolor{red}{\textbf{Case \# $1$}}: \textcolor{blue}{\textbf{Resistive Load as Linear Load with $R$}}}} & \multicolumn{3}{c|}{\textbf{Results}}\\ 
\cline{9-11}
\multicolumn{8}{|l||}{} & \multicolumn{2}{c|}{$\text{(THD)}_{\text{ANN}}$ [$\%$]} & \multirow{2}{*}{$\text{(THD)}_{\text{MPC}}$ [$\%$] ($t_{ss}$)}\\ 
\cline{1-10}
Sample No. & \multicolumn{2}{|c|}{$R$ [\SI{}{\ohm}]} & $T_s$ [\SI{}{\micro\second}] & $L$ [\SI{}{\milli\henry}] & $C$ [\SI{}{\micro\farad}] & $V_{dc}$ [\SI{}{\volt}] & $\text{v}^{*}_c$ [\SI{}{\volt}] & $\text{(THD)}_{S_1- S_{60}}$ & $\text{(THD)}_{S_1- S_{70}}$&\\ 
\hline \hline
$S_1$&\multicolumn{2}{|c|}{$10$}&$25$&$2.5$&$50$&$550$&$250$&\textbf{0.49}&$0.52$&$1.16$\;\;(\SI{2}{\milli\second})\\
$S_2$&\multicolumn{2}{|c|}{$30$}&$25$&$2.5$&$50$&$520$&$200$&\textbf{0.55}&$0.57$&$1.46$\;\;(\SI{5}{\milli\second})\\
$S_3$&\multicolumn{2}{|c|}{$50$}&$25$&$2.5$&$50$&$500$&$250$&\textbf{0.65}&$0.68$&$1.59$\;\;(\SI{5}{\milli\second})\\
$S_4$&\multicolumn{2}{|c|}{$80$}&$25$&$2.5$&$50$&$500$&$150$&\textbf{0.66}&$0.70$&$1.58$\;\;(\SI{10}{\milli\second})\\
$S_5$&\multicolumn{2}{|c|}{$300$}&$25$&$2.0$&$50$&$450$&$200$&\textbf{0.63}&$0.65$&$2.32$\;\;(\SI{20}{\milli\second})\\
$S_6$&\multicolumn{2}{|c|}{$500$}&$25$&$2.0$&$40$&$550$&$250$&\textbf{0.95}&$1.06$&$2.84$\;\;(\SI{35}{\milli\second})\\
$S_7$&\multicolumn{2}{|c|}{\SI{1}{\kilo\ohm}}&$25$&$3.5$&$40$&$520$&$200$&$0.72$&\textbf{0.70}&$1.51$\;\;(\SI{35}{\milli\second})\\
$S_8$&\multicolumn{2}{|c|}{\SI{2}{\mega\ohm}}&$25$&$4.0$&$40$&$500$&$150$&\textbf{0.76}&$0.84$&$1.31$\;\;(\SI{30}{\milli\second})\\
$S_9$&\multicolumn{2}{|c|}{\SI{10}{\mega\ohm}}&$25$&$2.0$&$40$&$500$&$200$&$0.99$&\textbf{0.98}&$2.61$\;\;(\SI{10}{\milli\second})\\
$S_{10}$&\multicolumn{2}{|c|}{Open Circuit}&$25$&$3.5$&$40$&$450$&$150$&\textbf{0.79}&$0.83$&$1.15$\;\;(\SI{30}{\milli\second})\\
\hdashline
$S_{11}$&\multicolumn{2}{|c|}{$15$}&$30$&$2.5$&$50$&$550$&$250$&\textbf{0.72}&$0.74$&$1.75$\\
$S_{12}$&\multicolumn{2}{|c|}{$40$}&$30$&$2.5$&$50$&$520$&$200$&$0.86$&\textbf{0.83}&$2.04$\\
$S_{13}$&\multicolumn{2}{|c|}{$100$}&$30$&$2.5$&$50$&$500$&$250$&\textbf{0.88}&$1.12$&$2.36$\\
$S_{14}$&\multicolumn{2}{|c|}{$200$}&$30$&$2.5$&$50$&$500$&$150$&$0.98$&\textbf{0.95}&$2.40$\\
$S_{15}$&\multicolumn{2}{|c|}{$300$}&$30$&$2.0$&$50$&$450$&$200$&\textbf{0.96}&$0.99$&$3.33$\\
$S_{16}$&\multicolumn{2}{|c|}{$500$}&$30$&$2.0$&$40$&$500$&$200$&\textbf{1.58}&$1.82$&$3.49$\\
$S_{17}$&\multicolumn{2}{|c|}{\SI{2}{\kilo\ohm}}&$30$&$3.5$&$40$&$520$&$200$&$1.15$&\textbf{1.09}&$2.36$\\
$S_{18}$&\multicolumn{2}{|c|}{\SI{1}{\mega\ohm}}&$30$&$4.0$&$40$&$500$&$150$&\textbf{1.20}&$1.27$&$1.88$\\
$S_{19}$&\multicolumn{2}{|c|}{\SI{5}{\mega\ohm}}&$30$&$2$&$40$&$500$&$200$&\textbf{1.61}&$1.62$&$3.91$\\
$S_{20}$&\multicolumn{2}{|c|}{Open Circuit}&$30$&$2.0$&$40$&$450$&$250$&\textbf{2.25}&$2.25$&$5.34$\\
\hdashline
$S_{21}$&\multicolumn{2}{|c|}{$20$}&$35$&$2.5$&$50$&$550$&$250$&\textbf{1.11}&$1.21$&$2.67$\\
$S_{22}$&\multicolumn{2}{|c|}{$100$}&$35$&$2.0$&$50$&$520$&$200$&\textbf{1.66}&$1.66$&$3.85$\\
$S_{23}$&\multicolumn{2}{|c|}{$250$}&$35$&$3.5$&$40$&$500$&$150$&\textbf{2.10}&$2.33$&$2.92$\\
$S_{24}$&\multicolumn{2}{|c|}{$400$}&$40$&$2.5$&$50$&$500$&$200$&\textbf{2.27}&$2.48$&$4.11$\\
$S_{25}$&\multicolumn{2}{|c|}{$500$}&$40$&$4.0$&$45$&$450$&$200$&\textbf{1.50}&$1.89$&$2.91$\\
\hdashline
$S_{26}$&\multicolumn{2}{|c|}{$400$}&$40$&$2.5$&$40$&$500$&$200$&$5.40$&$4.87$&\textbf{4.42}\\
$S_{27}$&\multicolumn{2}{|c|}{\SI{3}{\kilo\ohm}}&$35$&$2.0$&$40$&$550$&$200$&$4.23$&$4.30$&\textbf{4.19}\\
$S_{28}$&\multicolumn{2}{|c|}{\SI{3}{\kilo\ohm}}&$40$&$2.0$&$40$&$500$&$150$&$8.44$&$8.87$&\textbf{5.63}\\
$S_{29}$&\multicolumn{2}{|c|}{\SI{1}{\mega\ohm}}&$35$&$3.0$&$35$&$500$&$200$&$4.66$&$4.81$&\textbf{3.61}\\
$S_{30}$&\multicolumn{2}{|c|}{Open Circuit}&$40$&$3.5$&$40$&$450$&$150$&$5.39$&$5.15$&\textbf{3.70}\\
\hline\hline
\multicolumn{8}{|l||}{\multirow{2}{*}{\textcolor{red}{\textbf{Case \# $2$}}: \textcolor{blue}{\textbf{Diode-Bridge Rectifier as Non-Linear Load with $R_{NL}$ and $C_{NL}$}}}} & \multicolumn{3}{c|}{\textbf{Results}}\\
\cline{9-11}
\multicolumn{8}{|l||}{}& \multicolumn{2}{c|}{$\text{(THD)}_{\text{ANN}}$ [$\%$]} & \multirow{2}{*}{$\text{(THD)}_{\text{MPC}}$ [$\%$]}\\
\cline{1-10}
Sample No. & $R_{NL}$ [\SI{}{\ohm}] & $C_{NL}$ [\SI{}{\micro\farad}] & $T_s$ [\SI{}{\micro\second}] & $L$ [\SI{}{\milli\henry}] & $C$ [\SI{}{\micro\farad}] & $V_{dc}$ [\SI{}{\volt}] & $\text{v}^{*}_c$ [\SI{}{\volt}] & $\text{(THD)}_{S_1- S_{60}}$ & $\text{(THD)}_{S_1- S_{70}}$ & \\ 
\hline \hline
$S_{31}$&$10$&$3000$&$25$&$2.4$&$50$&$520$&$200$&\textbf{3.97}&$3.97$&$7.44$\\
$S_{32}$&$30$&$3000$&$25$&$3.5$&$50$&$500$&$200$&\textbf{1.99}&$2.10$&$3.80$\\
$S_{33}$&$60$&$3000$&$25$&$2.0$&$50$&$500$&$250$&\textbf{1.97}&$1.98$&$2.76$\\
$S_{34}$&\SI{1}{\kilo\ohm}&$3000$&$25$&$2.4$&$40$&$550$&$150$&$0.97$&\textbf{0.94}&$2.11$\\
$S_{35}$&\SI{1}{\kilo\ohm}&$200$&$25$&$3.5$&$35$&$520$&$200$&\textbf{0.80}&$0.90$&$1.40$\\
$S_{36}$&$60$&$100$&$25$&$4.0$&$40$&$450$&$150$&$1.36$&\textbf{1.30}&$1.64$\\
$S_{37}$&$100$&$1000$&$25$&$2.5$&$30$&$520$&$250$&\textbf{2.22}&$2.47$&$2.98$\\
\hdashline
$S_{38}$&$20$&$2000$&$33$&$3.0$&$50$&$520$&$200$&\textbf{2.94}&$3.10$&$4.57$\\
$S_{39}$&$30$&$2000$&$33$&$3.5$&$40$&$500$&$200$&$3.77$&\textbf{3.32}&$3.69$\\
$S_{40}$&$60$&$2000$&$33$&$2.0$&$50$&$500$&$250$&\textbf{3.15}&$3.31$&$4.08$\\
$S_{41}$&\SI{2}{\kilo\ohm}&$3000$&$33$&$2.4$&$40$&$550$&$150$&$2.90$&\textbf{2.89}&$3.65$\\
$S_{42}$&\SI{2}{\kilo\ohm}&$200$&$33$&$3.5$&$35$&$520$&$200$&$2.22$&\textbf{2.19}&$2.55$\\
$S_{43}$&$60$&$100$&$33$&$4.0$&$40$&$450$&$150$&\textbf{1.69}&$1.74$&$1.97$\\
$S_{44}$&$80$&$1000$&$33$&$4.0$&$35$&$520$&$250$&\textbf{3.47}&$3.60$&$3.66$\\
\hdashline
$S_{45}$&$100$&$3000$&$40$&$3.5$&$50$&$500$&$200$&\textbf{2.04}&$2.22$&$2.92$\\
$S_{46}$&$900$&$3000$&$40$&$3.0$&$40$&$520$&$250$&$4.32$&\textbf{4.30}&$4.65$\\
$S_{47}$&$100$&$1000$&$40$&$4.0$&$50$&$450$&$200$&$2.73$&\textbf{2.70}&$2.84$\\
$S_{48}$&$100$&$5000$&$40$&$4.0$&$45$&$520$&$250$&$3.71$&$3.79$&\textbf{3.63}\\
$S_{49}$&\SI{1}{\kilo\ohm}&$3000$&$40$&$2.5$&$\textbf{35}$&$500$&$150$&\color{red}{\textbf{22.23}}&\color{red}{\textbf{21.50}}&\textbf{5.78}\\
$S_{50}$&\SI{1}{\kilo\ohm}&$3000$&$40$&$2.5$&$\textbf{50}$&$500$&$150$&$2.41$&\textbf{2.30}&$4.76$\\
\hline
\end{tabular}
\label{table:Comparison}
\end{table*}

At the moment, one can say that the main limitation of the proposed method is that only the simulation results are not sufficient to prove its novelty in practical applications. However, indeed we believe that our proposed approach will also represent a novel contribution to the practical applications for the following reasons: 
(i) based on the previously proposed literature, both ANN-based and MPC-based approaches have shown good results in both simulated and experimental scenarios;
(ii) moreover, the trained network is only required to be fine-tuned, in order to improve its performance in practical applications.

\section{Conclusions and Future Work}\label{CONCLUSION}
A novel control strategy using a feed-forward ANN to generate a high-quality sinusoidal output voltage of a three-phase inverter with an output $LC$ filter has been successfully developed and tested, for different types of loads under various operating conditions. The output voltage of the inverter is directly controlled, without the need for the mathematical model of the inverter, considering the whole system as a black-box. In this work, MPC has been used for two main purposes: (i) generating the data required for the off-line training of the proposed ANN, and (ii) comparing its performance with the proposed ANN-based controller for linear and non-linear load conditions. Simulation results, based on fifty test 
different than those that were used at training time, show that the proposed ANN-based controller performs better than MPC, in terms of a lower THD and a fast and safe transient response, demonstrating the excellent steady and dynamic performance of the proposed ANN-based control strategy. As in any model-based control strategy, variations in the system parameters inevitably influence the performance of the ANN-based control scheme proposed in this paper. 
The possible directions for future work would be (i) the implementation of the ANN-based controller in practical applications; then (ii) the employment in other power electronics applications, possibly employing different neural networks.

\bibliographystyle{IEEEtran}
\bibliography{References}

\begin{thebibliography}{10}
\providecommand{\url}[1]{#1}
\csname url@samestyle\endcsname
\providecommand{\newblock}{\relax}
\providecommand{\bibinfo}[2]{#2}
\providecommand{\BIBentrySTDinterwordspacing}{\spaceskip=0pt\relax}
\providecommand{\BIBentryALTinterwordstretchfactor}{4}
\providecommand{\BIBentryALTinterwordspacing}{\spaceskip=\fontdimen2\font plus
\BIBentryALTinterwordstretchfactor\fontdimen3\font minus
  \fontdimen4\font\relax}
\providecommand{\BIBforeignlanguage}[2]{{%
\expandafter\ifx\csname l@#1\endcsname\relax
\typeout{** WARNING: IEEEtran.bst: No hyphenation pattern has been}%
\typeout{** loaded for the language `#1'. Using the pattern for}%
\typeout{** the default language instead.}%
\else
\language=\csname l@#1\endcsname
\fi
#2}}
\providecommand{\BIBdecl}{\relax}
\BIBdecl

\bibitem{cortes2009model}
P.~Cort{\'e}s, G.~Ortiz, J.~I. Yuz, J.~Rodr{\'\i}guez, S.~Vazquez, and L.~G.
  Franquelo, ``Model predictive control of an inverter with output {LC} filter
  for {UPS} applications,'' \emph{IEEE Transactions on Industrial Electronics},
  vol.~56, no.~6, pp. 1875--1883, 2009.

\bibitem{habetler2002design}
T.~G. Habetler, R.~Naik, and T.~A. Nondahl, ``Design and implementation of an
  inverter output {LC} filter used for dv/dt reduction,'' \emph{IEEE
  Transactions on Power Electronics}, vol.~17, no.~3, pp. 327--331, 2002.

\bibitem{kazmierkowski1998current}
M.~P. Kazmierkowski and L.~Malesani, ``Current control techniques for
  three-phase voltage-source {PWM} converters: {A} survey,'' \emph{IEEE
  Transactions on industrial electronics}, vol.~45, no.~5, pp. 691--703, 1998.

\bibitem{carrasco2006power}
J.~M. Carrasco, L.~G. Franquelo, J.~T. Bialasiewicz, E.~Galv{\'a}n, R.~C.
  PortilloGuisado, M.~M. Prats, J.~I. Le{\'o}n, and N.~Moreno-Alfonso,
  ``Power-electronic systems for the grid integration of renewable energy
  sources: {A} survey,'' \emph{IEEE Transactions on industrial electronics},
  vol.~53, no.~4, pp. 1002--1016, 2006.

\bibitem{blaabjerg2006overview}
F.~Blaabjerg, R.~Teodorescu, M.~Liserre, and A.~V. Timbus, ``Overview of
  control and grid synchronization for distributed power generation systems,''
  \emph{IEEE Transactions on industrial electronics}, vol.~53, no.~5, pp.
  1398--1409, 2006.

\bibitem{gurrero2007uninterruptible}
J.~Gurrero, L.~G. De~Vicuna, and J.~Uceda, ``Uninterruptible power supply
  systems provide protection,'' \emph{IEEE Industrial Electronics Magazine},
  vol.~1, no.~1, pp. 28--38, 2007.

\bibitem{hung1993variable}
J.~Y. Hung, W.~Gao, and J.~C. Hung, ``Variable structure control: {A} survey,''
  \emph{IEEE transactions on industrial electronics}, vol.~40, no.~1, pp.
  2--22, 1993.

\bibitem{mohamed2013classical}
I.~S. Mohamed, S.~A. Zaid, M.~Abu-Elyazeed, and H.~M. Elsayed, ``Classical
  methods and model predictive control of three-phase inverter with output {LC}
  filter for {UPS} applications,'' in \emph{Control, Decision and Information
  Technologies (CoDIT), 2013 International Conference on}.\hskip 1em plus 0.5em
  minus 0.4em\relax IEEE, 2013, pp. 483--488.

\bibitem{brod1985current}
D.~M. Brod and D.~W. Novotny, ``Current control of {VSI-PWM} inverters,''
  \emph{IEEE Transactions on Industry Applications}, no.~3, pp. 562--570, 1985.

\bibitem{jung2004optimal}
J.~Jung, M.~Dai, and A.~Keyhani, ``Optimal control of three-phase {PWM}
  inverter for {UPS} systems,'' in \emph{IEEE Power Electronics Specialists
  Conference}, vol.~3, 2004, pp. 2054--2059.

\bibitem{mohamed2013model}
I.~S. Mohamed, S.~A. Zaid, M.~Abu-Elyazeed, and H.~M. Elsayed, ``Model
  predictive control-a simple and powerful method to control {UPS} inverter
  applications with output {LC} filter,'' in \emph{Electronics, Communications
  and Photonics Conference (SIECPC), 2013 Saudi International}.\hskip 1em plus
  0.5em minus 0.4em\relax IEEE, 2013, pp. 1--6.

\bibitem{rojas2017new}
F.~Rojas, R.~Kennel, R.~Cardenas, R.~Repenning, J.~C. Clare, and M.~Diaz, ``A
  new space-vector-modulation algorithm for a three-level four-leg {NPC}
  inverter,'' \emph{IEEE Transactions on Energy Conversion}, vol.~32, no.~1,
  pp. 23--35, 2017.

\bibitem{loh2003comparative}
P.~C. Loh, M.~J. Newman, D.~N. Zmood, and D.~G. Holmes, ``A comparative
  analysis of multiloop voltage regulation strategies for single and
  three-phase {UPS} systems,'' \emph{IEEE Transactions on Power Electronics},
  vol.~18, no.~5, pp. 1176--1185, 2003.

\bibitem{loh2005analysis}
P.~C. Loh and D.~G. Holmes, ``Analysis of multiloop control strategies for
  {LC/CL/LCL}-filtered voltage-source and current-source inverters,''
  \emph{IEEE Transactions on Industry Applications}, vol.~41, no.~2, pp.
  644--654, 2005.

\bibitem{mohamed2007improved}
Y.~A.-R.~I. Mohamed and E.~F. El-Saadany, ``An improved deadbeat current
  control scheme with a novel adaptive self-tuning load model for a three-phase
  {PWM} voltage-source inverter,'' \emph{IEEE Transactions on Industrial
  Electronics}, vol.~54, no.~2, pp. 747--759, 2007.

\bibitem{lim2014robust}
J.~S. Lim, C.~Park, J.~Han, and Y.~I. Lee, ``Robust tracking control of a
  three-phase {DC--AC} inverter for {UPS} applications,'' \emph{IEEE
  Transactions on Industrial Electronics}, vol.~61, no.~8, pp. 4142--4151,
  2014.

\bibitem{pichan2017deadbeat}
M.~Pichan, H.~Rastegar, and M.~Monfared, ``Deadbeat control of the stand-alone
  four-leg inverter considering the effect of the neutral line inductor,''
  \emph{IEEE Trans. Industrial Electronics}, vol.~64, no.~4, pp. 2592--2601,
  2017.

\bibitem{escobar2007adaptive}
G.~Escobar, P.~Mattavelli, A.~M. Stankovic, A.~A. Valdez, and J.~Leyva-Ramos,
  ``An adaptive control for {UPS} to compensate unbalance and harmonic
  distortion using a combined capacitor/load current sensing,'' \emph{IEEE
  Transactions on Industrial Electronics}, vol.~54, no.~2, pp. 839--847, 2007.

\bibitem{jiang2012low}
S.~Jiang, D.~Cao, Y.~Li, J.~Liu, and F.~Z. Peng, ``{Low-THD}, fast-transient,
  and cost-effective synchronous-frame repetitive controller for three-phase
  {UPS} inverters,'' \emph{IEEE Transactions on Power Electronics}, vol.~27,
  no.~6, pp. 2994--3005, 2012.

\bibitem{wu2005digital}
E.~Wu and P.~W. Lehn, ``Digital current control of a voltage source converter
  with active damping of {LCL} resonance,'' in \emph{Twentieth Annual IEEE
  Applied Power Electronics Conference and Exposition, 2005. APEC 2005.},
  vol.~3.\hskip 1em plus 0.5em minus 0.4em\relax IEEE, 2005, pp. 1642--1649.

\bibitem{komurcugil2012rotating}
H.~Komurcugil, ``Rotating-sliding-line-based sliding-mode control for
  single-phase {UPS} inverters,'' \emph{IEEE Transactions on Industrial
  Electronics}, vol.~59, no.~10, pp. 3719--3726, 2012.

\bibitem{sabir2017robust}
S.~Sabir, Q.~Khan, M.~Saleem, and A.~Khaliq, ``Robust voltage tracking control
  of three phase inverter with an output {LC} filter: {A} sliding mode
  approach,'' in \emph{Emerging Technologies (ICET), 2017 13th International
  Conference on}.\hskip 1em plus 0.5em minus 0.4em\relax IEEE, 2017, pp. 1--5.

\bibitem{cortes2008predictive}
P.~Cort{\'e}s, M.~P. Kazmierkowski, R.~Kennel, D.~E. Quevedo, and J.~R.
  Rodriguez, ``Predictive control in power electronics and drives.'' \emph{IEEE
  Trans. Industrial Electronics}, vol.~55, no.~12, pp. 4312--4324, 2008.

\bibitem{singh2018hil}
V.~K. Singh, R.~N. Tripathi, and T.~Hanamoto, ``{HIL} co-simulation of finite
  set-model predictive control using {FPGA} for a three-phase {VSI} system,''
  \emph{Energies}, vol.~11, no.~4, p. 909, 2018.

\bibitem{willmann2007multiple}
G.~Willmann, D.~F. Coutinho, L.~F.~A. Pereira, and F.~B. L{\'\i}bano,
  ``Multiple-loop {H-infinity} control design for uninterruptible power
  supplies,'' \emph{IEEE Transactions on Industrial Electronics}, vol.~54,
  no.~3, pp. 1591--1602, 2007.

\bibitem{lee2004robust}
T.-S. Lee, K.~Tzeng, and M.~Chong, ``Robust controller design for a
  single-phase {UPS} inverter using $\mu$-synthesis,'' \emph{IEE
  Proceedings-Electric Power Applications}, vol. 151, no.~3, pp. 334--340,
  2004.

\bibitem{nauman2016efficient}
M.~Nauman and A.~Hasan, ``Efficient implicit model-predictive control of a
  three-phase inverter with an output {LC} filter,'' \emph{IEEE Transactions on
  Power Electronics}, vol.~31, no.~9, pp. 6075--6078, 2016.

\bibitem{mohamed2016implementation}
I.~S. Mohamed, S.~A. Zaid, M.~Abu-Elyazeed, and H.~M. Elsayed, ``Implementation
  of model predictive control for three-phase inverter with output {LC} filter
  on {eZdsp} {F28335 Kit} using {HIL} simulation,'' \emph{International Journal
  of Modelling, Identification and Control}, vol.~25, no.~4, pp. 301--312,
  2016.

\bibitem{guo2019improved}
L.~Guo, N.~Jin, C.~Gan, L.~Xu, and Q.~Wang, ``An improved model predictive
  control strategy to reduce common-mode voltage for two-level voltage source
  inverters considering dead-time effects,'' \emph{IEEE Transactions on
  Industrial Electronics}, vol.~66, no.~5, pp. 3561--3572, 2019.

\bibitem{camacho2007nonlinear}
E.~F. Camacho and C.~Bordons, ``Nonlinear model predictive control: {An}
  introductory review,'' in \emph{Assessment and future directions of nonlinear
  model predictive control}.\hskip 1em plus 0.5em minus 0.4em\relax Springer,
  2007, pp. 1--16.

\bibitem{vazquez2017model}
S.~Vazquez, J.~Rodriguez, M.~Rivera, L.~G. Franquelo, and M.~Norambuena,
  ``Model predictive control for power converters and drives: Advances and
  trends,'' \emph{IEEE Transactions on Industrial Electronics}, vol.~64, no.~2,
  pp. 935--947, 2017.

\bibitem{nguyen2017model}
H.~T. Nguyen, E.-K. Kim, I.-P. Kim, H.~H. Choi, and J.-W. Jung, ``Model
  predictive control with modulated optimal vector for a three-phase inverter
  with an {LC} filter,'' \emph{IEEE Transactions on Power Electronics},
  vol.~33, no.~3, pp. 2690--2703, 2017.

\bibitem{vazquez2017fcs}
S.~Vazquez, A.~Marquez, J.~I. Leon, L.~G. Franquelo, and T.~Geyer, ``{FCS-MPC}
  and observer design for a {VSI} with output {LC} filter and sinusoidal output
  currents,'' in \emph{2017 11th IEEE International Conference on
  Compatibility, Power Electronics and Power Engineering (CPE-POWERENG)}.\hskip
  1em plus 0.5em minus 0.4em\relax IEEE, 2017, pp. 677--682.

\bibitem{mohamed2015improved}
I.~S. Mohamed, S.~A. Zaid, M.~Abu-Elyazeed, and H.~M. Elsayed, ``Improved model
  predictive control for three-phase inverter with output {LC} filter,''
  \emph{International Journal of Modelling, Identification and Control},
  vol.~23, no.~4, pp. 371--379, 2015.

\bibitem{dragivcevic2017model}
T.~Dragi{\v{c}}evi{\'c}, ``Model predictive control of power converters for
  robust and fast operation of {AC} microgrids,'' \emph{IEEE Transactions on
  Power Electronics}, vol.~33, no.~7, pp. 6304--6317, 2017.

\bibitem{zheng2019current}
C.~Zheng, T.~Dragicevic, and F.~Blaabjerg, ``Current-sensorless finite-set
  model predictive control for {LC}-filtered voltage source inverters,''
  \emph{IEEE Transactions on Power Electronics}, 2019.

\bibitem{mariethoz2009explicit}
S.~Mari{\'e}thoz and M.~Morari, ``Explicit model-predictive control of a {PWM}
  inverter with an {LCL} filter,'' \emph{IEEE Transactions on Industrial
  Electronics}, vol.~56, no.~2, pp. 389--399, 2009.

\bibitem{kwak2014switching}
S.~Kwak and J.-C. Park, ``Switching strategy based on model predictive control
  of {VSI} to obtain high efficiency and balanced loss distribution,''
  \emph{IEEE Trans. Power Electron}, vol.~29, no.~9, pp. 4551--4567, 2014.

\bibitem{rashid2017power}
M.~H. Rashid, \emph{Power electronics handbook}.\hskip 1em plus 0.5em minus
  0.4em\relax Butterworth-Heinemann, 2017.

\bibitem{narendra1990identification}
K.~S. Narendra and K.~Parthasarathy, ``Identification and control of dynamical
  systems using neural networks,'' \emph{IEEE Transactions on neural networks},
  vol.~1, no.~1, pp. 4--27, 1990.

\bibitem{hunt1992neural}
K.~J. Hunt, D.~Sbarbaro, R.~{\.Z}bikowski, and P.~J. Gawthrop, ``Neural
  networks for control systems-a survey,'' \emph{Automatica}, vol.~28, no.~6,
  pp. 1083--1112, 1992.

\bibitem{saint1991neural}
J.~Saint-Donat, N.~Bhat, and T.~J. McAvoy, ``Neural net based model predictive
  control,'' \emph{International Journal of Control}, vol.~54, no.~6, pp.
  1453--1468, 1991.

\bibitem{karanayil2018artificial}
B.~Karanayil and M.~F. Rahman, ``Artificial neural network applications in
  power electronics and electric drives,'' in \emph{Power Electronics Handbook
  (Fourth Edition)}.\hskip 1em plus 0.5em minus 0.4em\relax Elsevier, 2018, pp.
  1245--1260.

\bibitem{wishart1995identification}
M.~T. Wishart and R.~G. Harley, ``Identification and control of induction
  machines using artificial neural networks,'' \emph{IEEE Transactions on
  Industry Applications}, vol.~31, no.~3, pp. 612--619, 1995.

\bibitem{sun2013speed}
X.~Sun, L.~Chen, Z.~Yang, and H.~Zhu, ``Speed-sensorless vector control of a
  bearingless induction motor with artificial neural network inverse speed
  observer,'' \emph{IEEE/ASME Transactions on mechatronics}, vol.~18, no.~4,
  pp. 1357--1366, 2013.

\bibitem{lee2018performance}
H.-Y. Lee, J.-L. Lee, S.-O. Kwon, and S.-W. Lee, ``Performance estimation of
  induction motor using artificial neural network,'' in \emph{2018 25th
  International Conference on Systems, Signals and Image Processing
  (IWSSIP)}.\hskip 1em plus 0.5em minus 0.4em\relax IEEE, 2018, pp. 1--3.

\bibitem{gadoue2013stator}
S.~M. Gadoue, D.~Giaouris, and J.~W. Finch, ``Stator current model reference
  adaptive systems speed estimator for regenerating-mode low-speed operation of
  sensorless induction motor drives,'' \emph{IET Electric Power Applications},
  vol.~7, no.~7, pp. 597--606, 2013.

\bibitem{bakhshai1996combined}
A.~Bakhshai, J.~Espinoza, G.~Joos, and H.~Jin, ``A combined artificial neural
  network and {DSP} approach to the implementation of space vector modulation
  techniques,'' in \emph{Industry Applications Conference, 1996. Thirty-First
  IAS Annual Meeting, IAS'96., Conference Record of the 1996 IEEE},
  vol.~2.\hskip 1em plus 0.5em minus 0.4em\relax IEEE, 1996, pp. 934--940.

\bibitem{pinto2000neural}
J.~O. Pinto, B.~K. Bose, L.~B. Da~Silva, and M.~P. Kazmierkowski, ``A
  neural-network-based space-vector {PWM} controller for voltage-fed inverter
  induction motor drive,'' \emph{IEEE Transactions on Industry Applications},
  vol.~36, no.~6, pp. 1628--1636, 2000.

\bibitem{karatepe2009artificial}
E.~Karatepe, T.~Hiyama \emph{et~al.}, ``Artificial neural network-polar
  coordinated fuzzy controller based maximum power point tracking control under
  partially shaded conditions,'' \emph{IET Renewable Power Generation}, vol.~3,
  no.~2, pp. 239--253, 2009.

\bibitem{akter2016modified}
M.~P. Akter, S.~Mekhilef, N.~M.~L. Tan, and H.~Akagi, ``Modified model
  predictive control of a bidirectional {AC--DC} converter based on lyapunov
  function for energy storage systems,'' \emph{IEEE Transactions on Industrial
  Electronics}, vol.~63, no.~2, pp. 704--715, 2016.

\bibitem{lin1993power}
B.-R. Lin and R.~G. Hoft, ``Power electronics inverter control with neural
  networks,'' in \emph{Proceedings Eighth Annual Applied Power Electronics
  Conference and Exposition,}.\hskip 1em plus 0.5em minus 0.4em\relax IEEE,
  1993, pp. 128--134.

\bibitem{sun2002analogue}
X.~Sun, M.~H. Chow, F.~H. Leung, D.~Xu, Y.~Wang, and Y.-S. Lee, ``Analogue
  implementation of a neural network controller for {UPS} inverter
  applications,'' \emph{IEEE transactions on power electronics}, vol.~17,
  no.~3, pp. 305--313, 2002.

\bibitem{boumaaraf2015three}
H.~Boumaaraf, A.~Talha, and O.~Bouhali, ``A three-phase {NPC} grid-connected
  inverter for photovoltaic applications using neural network {MPPT},''
  \emph{Renewable and Sustainable Energy Reviews}, vol.~49, pp. 1171--1179,
  2015.

\bibitem{wai2015design}
R.-J. Wai, M.-W. Chen, and Y.-K. Liu, ``Design of adaptive control and fuzzy
  neural network control for single-stage boost inverter,'' \emph{IEEE
  Transactions on Industrial Electronics}, vol.~62, no.~9, pp. 5434--5445,
  2015.

\bibitem{fu2016control}
X.~Fu and S.~Li, ``Control of single-phase grid-connected converters with {LCL}
  filters using recurrent neural network and conventional control methods,''
  \emph{IEEE Transactions on Power Electronics}, vol.~31, no.~7, pp.
  5354--5364, 2016.

\bibitem{mohamed2013three}
I.~S. Mohamed, S.~A. Zaid, H.~M. Elsayed, and M.~Abu-Elyazeed, ``Three-phase
  inverter with output {LC} filter using predictive control for {UPS}
  applications,'' in \emph{Control, Decision and Information Technologies
  (CoDIT), 2013 International Conference on}.\hskip 1em plus 0.5em minus
  0.4em\relax IEEE, 2013, pp. 489--494.

\bibitem{piche2000nonlinear}
S.~Piche, B.~Sayyar-Rodsari, D.~Johnson, and M.~Gerules, ``Nonlinear model
  predictive control using neural networks,'' \emph{IEEE Control Systems
  Magazine}, vol.~20, no.~3, pp. 53--62, 2000.

\bibitem{aakesson2006neural}
B.~M. {\AA}kesson and H.~T. Toivonen, ``A neural network model predictive
  controller,'' \emph{Journal of Process Control}, vol.~16, no.~9, pp.
  937--946, 2006.

\bibitem{wang2018deep}
D.~Wang, X.~Yin, S.~Tang, C.~Zhang, Z.~J. Shen, J.~Wang, and Z.~Shuai, ``A deep
  neural network based predictive control strategy for high frequency
  multilevel converters,'' in \emph{2018 IEEE Energy Conversion Congress and
  Exposition (ECCE)}.\hskip 1em plus 0.5em minus 0.4em\relax IEEE, 2018, pp.
  2988--2992.

\bibitem{dragicevic2018weighting}
T.~Dragicevic and M.~Novak, ``Weighting factor design in model predictive
  control of power electronic converters: {An} artificial neural network
  approach,'' \emph{IEEE Transactions on Industrial Electronics}, 2018.

\bibitem{hornik1991approximation}
K.~Hornik, ``Approximation capabilities of multilayer feedforward networks,''
  \emph{Neural networks}, vol.~4, no.~2, pp. 251--257, 1991.

\bibitem{FP16:2019}
\emph{Training with Mixed precision}, NVIDIA, June 2019,
  \url{https://docs.nvidia.com/deeplearning/sdk/pdf/Training-Mixed-Precision-User-Guide.pdf}.

\bibitem{Int16:2018}
D.~Das, N.~Mellempudi, D.~Mudigere, D.~Kalamkar, S.~Avancha, K.~Banerjee,
  S.~Sridharan, K.~Vaidyanathan, B.~Kaul, E.~Georganas, A.~Heinecke, P.~Dubey,
  J.~Corbal, N.~Shustrov, R.~Dubtsov, E.~Fomenko, and V.~Pirogov, ``Mixed
  precision training of convolutional neural networks using integer
  operations,'' no. arXiv:1802.00930 [cs.NE], 2018.

\bibitem{Hu:2019}
Z.~Hu, A.~B. Tarakji, V.~Raheja, C.~Phillips, T.~Wang, and I.~Mohomed,
  ``Deephome: Distributed inference with heterogeneous devices in the edge,''
  in \emph{The 3rd International Workshop on Deep Learning for Mobile Systems
  and Applications}, ser. EMDL '19.\hskip 1em plus 0.5em minus 0.4em\relax New
  York, NY, USA: ACM, 2019, pp. 13--18.

\bibitem{Moller1993}
M.~F. M{\o}ller, ``A scaled conjugate gradient algorithm for fast supervised
  learning,'' \emph{Neural networks}, vol.~6, no.~4, pp. 525--533, 1993.

\bibitem{Fletcher2000}
R.~Fletcher, \emph{Practical methods of optimization}.\hskip 1em plus 0.5em
  minus 0.4em\relax John Wiley \& Sons, Ltd, 2000, second edition.

\end{thebibliography}
\begin{IEEEbiography}[{\includegraphics[width=1.1in,height=1.25in,clip,keepaspectratio]{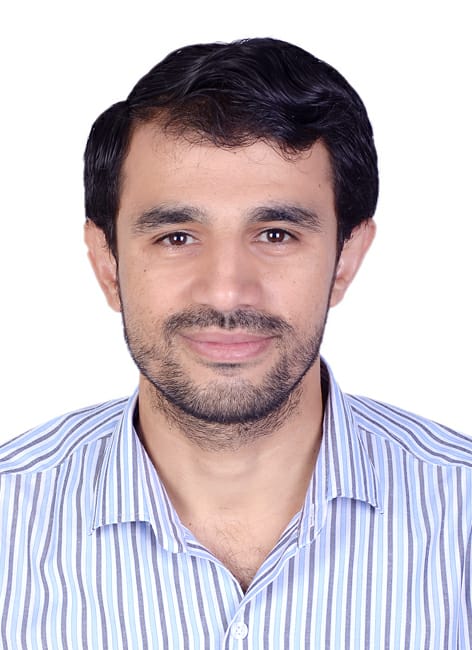}}]{Ihab S. Mohamed} received his B.S. degree from the Institute of Aviation Engineering and Technology (IAET) in 2009, Egypt. Afterwards, he received the M.S. degree in Electrical Engineering from Cairo University, Egypt, in 2014. From 2009 to 2015, he was appointed as a Teaching Assistant at Electronics and Communications Department, IAET.

From 2015 to 2017, he was an MSc student of ``European Master in Advanced Robotics (EMARO+)''. He attended the first year of EMARO+ at Warsaw University of Technology (WUT), Poland, while the second year was at University of Genoa (GU), Italy. He is currently pursuing a Ph.D. degree in Robotics Engineering at INRIA Sophia Antipolis - M\'{e}diterran\'{e}e, Universit\'{e} C\^{o}te d'Azur, France, under the supervision of Prof. Philippe Martinet, Guillaume Allibert, and Paolo Salaris. Ihab's research interests include predictive control (MPC), power electronics, robotics, computer vision, and machine learning. 
\end{IEEEbiography}
\begin{IEEEbiography}[{\includegraphics[width=1in,height=1.25in,clip,keepaspectratio]{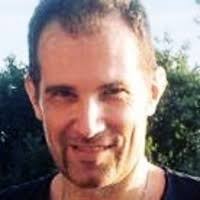}}]{Stefano Rovetta} (M'99-SM'12) is Associate Professor of Computer Science at the University of Genova (Italy). He authored more than 170 scientific papers in machine learning, neural networks, clustering, fuzzy systems and bioinformatics. 

He received the 2008 Pattern Recognition Society Award, and was the chair of international conferences. He is a member of the Italian Neural Network Society, the European Neural Network Society, and the European Society for Fuzzy Logic And Technology. 
\end{IEEEbiography}
\begin{IEEEbiography}[{\includegraphics[width=1in,height=1.25in,clip,keepaspectratio]{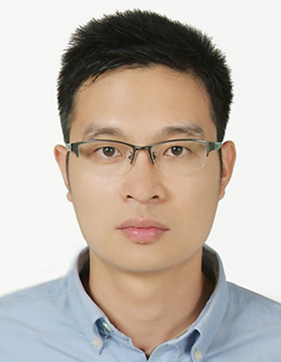}}]{Ton Duc Do} (S'12-M'14) received the B.S. and M.S. degrees in electrical engineering from Hanoi University of Science and Technology, Hanoi, Vietnam, in 2007 and 2009, respectively, and the Ph.D. Degree in electrical engineering from Dongguk University, Seoul, Korea, in 2014. 

From 2008 to 2009, he worked at the Division of Electrical Engineering, Thuy Loi University, Vietnam, as a Lecturer. He was at the Division of Electronics and Electrical Engineering, Dongguk University, as a Postdoctoral Researcher in 2014. He was also a senior researcher at the Pioneer Research Center for Controlling Dementia by Converging Technology, Gyeongsang National University, Korea from May 2014 to Aug. 2015. From Sep. 2015, he has been an assistant professor in the department of Robotics and Mechatronics, Nazarbayev University, Kazakhstan. His research interests include the field of advanced control system theories, electric machine drives, renewable energy conversion systems, uninterruptible power supplies, electromagnetic actuator systems, targeted drug delivery systems, and nanorobots. 

Dr. Do received the best research award from Dongguk University in 2014. He has been a member of IEEE since 2012. He was also the Lead Guest Editor for special issue for the special issue of Mathematical Problems in Engineering on ``Advanced Control Methods for Systems with Fast-Varying Disturbances and Applications''. He is currently an associate editor of IEEE Access. 
\end{IEEEbiography}
\begin{IEEEbiography}[{\includegraphics[width=1.0in,height=1.8in,clip,keepaspectratio]{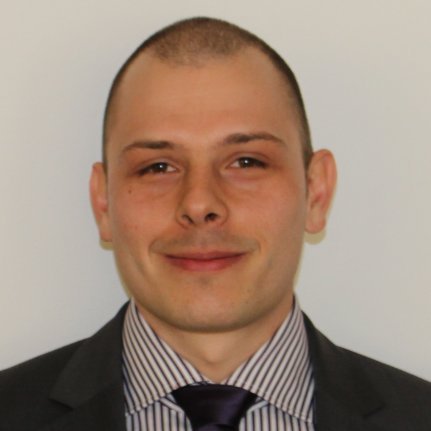}}]{Tomislav Dragi\u{c}evi\'{c}} (S'09-M'13-SM'17) received the M.Sc. and the industrial Ph.D. degrees in Electrical Engineering from the Faculty of Electrical Engineering, Zagreb, Croatia, in 2009 and 2013, respectively. From 2013 until 2016, he has been a Postdoctoral research associate at Aalborg University, Denmark. From March 2016, he is an Associate Professor at Aalborg University, Denmark where he leads an Advanced Control Lab. 

He made a guest professor stay at Nottingham University, UK during spring/summer of 2018. His principal field of interest is design and control of microgrids, and application of advanced modeling and control concepts to power electronic systems. He has authored and co-authored more than 170 technical papers (more than 70 of them are published in international journals, mostly IEEE Transactions) in his domain of interest, 8 book chapters and a book in the field.

He serves as Associate Editor in the IEEE TRANSACTIONS ON INDUSTRIAL ELECTRONICS, in IEEE Emerging and Selected Topics in Power Electronics and in IEEE Industrial Electronics Magazine. Dr. Dragi\u{c}evi\'{c} is a recipient of the Kon\u{c}ar prize for the best industrial PhD thesis in Croatia, and a Robert Mayer Energy Conservation award.
\end{IEEEbiography}
\begin{IEEEbiography}[{\includegraphics[width=1in,height=1.25in,clip,keepaspectratio]{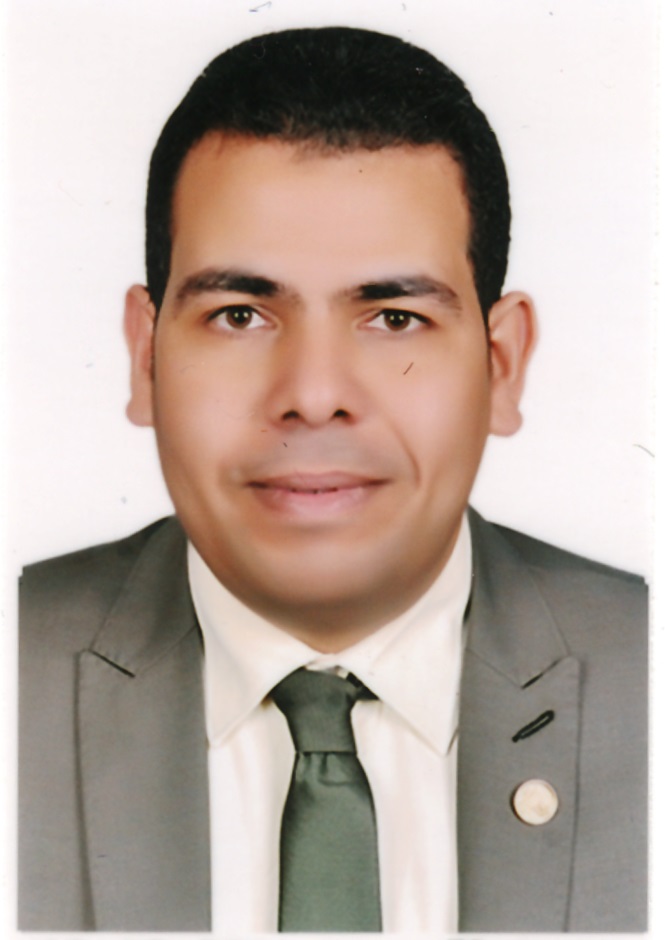}}]{Ahmed A. Zaki Diab} received B.Sc. and M.Sc. in Electrical Engineering from Minia University, Egypt in 2006 and 2009, respectively. In 2015, he received his PhD from Electric Drives and Industry Automation Department, Faculty of Mechatronics and Automation at Novosibirsk State Technical University, Novosibirsk, Russia. He had obtained postdoctoral Fellowship at the National research university ``MPEI'', Moscow Power Engineering Institute, Moscow, Russia from September 2017 to March 2018. Since 2001, he has been with the Department of Electrical Engineering, Faculty of Engineering, Minia University, Egypt as a Teaching Assistant, a Lecturer Assistant, and since 2015, as an Assistant Professor. His present research interests include Ac drives, application of control techniques and optimization algorithms in renewable energy systems.
\end{IEEEbiography}

\end{document}